\documentclass[sn-mathphys-num]{sn-jnl}

\usepackage{graphicx}%
\usepackage{multirow}%
\usepackage{amsmath,amssymb,amsfonts}%
\usepackage{amsthm}
\usepackage{mathrsfs}%
\usepackage[title]{appendix}%

\usepackage{textcomp}%
\usepackage{manyfoot}%
\usepackage{booktabs}%
\usepackage{algorithm}%
\usepackage{algorithmicx}%
\usepackage{algpseudocode}%
\usepackage{listings}%

\usepackage{lettrine}

\usepackage{lmodern}
\usepackage{anyfontsize}
\usepackage{epsfig}
\usepackage{color}
\usepackage{verbatim}
\usepackage{multirow}
\usepackage{color}
\usepackage{colortbl,hhline}
\usepackage{xcolor}
\usepackage{lmodern}
\usepackage{nomencl}
\usepackage{longtable}
\usepackage{caption}

\usepackage{lipsum}    

\usepackage{ltcaption} 
\usepackage{multicol}  
\makenomenclature
\usepackage{array}
\usepackage{diagbox}
\usepackage{arydshln}
\usepackage{stfloats}
\usepackage[utf8]{inputenc}
\usepackage[T1]{fontenc}
\usepackage{booktabs}
\usepackage{multirow}

\theoremstyle{thmstyleone}%
\theoremstyle{thmstyletwo}%

\theoremstyle{thmstylethree}%

\begin{document}

\title[Deep Learning for Personalized Electrocardiogram Diagnosis: A Review]{Deep Learning for Personalized Electrocardiogram Diagnosis: A Review}


\author[1]{\fnm{Cheng} \sur{Ding}}\email{chengding@gatech.edu}
\equalcont{These authors contributed equally to this work.}

\author[2]{\fnm{Tianliang} \sur{Yao}}\email{2150248@tongji.edu.cn}
\equalcont{These authors contributed equally to this work.}

\author[3]{\fnm{Chenwei} \sur{Wu}}\email{cwu59@u.rochester.edu}

\author*[4]{\fnm{Jianyuan} \sur{Ni}}\email{jni100@juniata.edu}

\affil[1]{\orgdiv{Department of Biomedical Engineering}, \orgname{Georgia Institution of Technlogy}, \orgaddress{\city{Atlanta}, \country{United States}}}

\affil[2]{\orgdiv{Department of Control Science and Engineering, College of Electronics and Information Engineering}, \orgname{Tongji University}, \orgaddress{\street{No. 1239, Siping Road},  \postcode{200092}, \state{Shanghai}, \country{China}}}

\affil[3]{\orgdiv{Department of Biomedical Engineering}, \orgname{University of Rochester}, \orgaddress{\city{Rochester}, \country{United States}}}

\affil*[4]{\orgdiv{Department of Information Technology and Computer Science}, \orgname{Juniata College}, \orgaddress{\city{Huntingdon}, \country{United States}}}


\abstract{The electrocardiogram (ECG) remains a fundamental tool in cardiac diagnostics, yet its interpretation traditionally reliant on the expertise of cardiologists. The emergence of deep learning has heralded a revolutionary era in medical data analysis, particularly in the domain of ECG diagnostics. However, inter-patient variability prohibit the generalibility of ECG-AI model trained on a population dataset, hence degrade the performance of ECG-AI on specific patient or patient group. Many studies have address this challenge using different deep learning technologies. This comprehensive review systematically synthesizes research from a wide range of studies to provide an in-depth examination of cutting-edge deep-learning techniques in personalized ECG diagnosis. The review outlines a rigorous methodology for the selection of pertinent scholarly articles and offers a comprehensive overview of deep learning approaches applied to personalized ECG diagnostics. Moreover, the challenges these methods encounter are investigated, along with future research directions, culminating in insights into how the integration of deep learning can transform personalized ECG diagnosis and enhance cardiac care. By emphasizing both the strengths and limitations of current methodologies, this review underscores the immense potential of deep learning to refine and redefine ECG analysis in clinical practice, paving the way for more accurate, efficient, and personalized cardiac diagnostics.}

\keywords{Personalized diagnosis, electrocardiogram (ECG), deep-learning, fine-tuning, transfer learning, generative adversarial networks(GANs), diffusion models, meta-learning, domain adaptation, inter-individual variability, healthcare data privacy.}



\maketitle

\section{Introduction}

The electrocardiogram (ECG) remains a cornerstone in the non-invasive assessment of cardiac electrophysiological activity, providing critical insights into myocardial function and overall cardiac health. This diagnostic tool captures the heart's electrical activity, facilitating the diagnosis of a spectrum of cardiovascular diseases, including arrhythmias, myocardial infarctions, and conduction system disorders \cite{bhatia2018screening,martis2014current,liu2021deep}. Traditional interpretation of ECG data, however, depends heavily on the acumen of cardiologists. This process is not only time-intensive but also suffers from subjective biases and considerable inter-observer variability, which may impact diagnostic accuracy \cite{rafie2021ecg}.

With the advent of deep learning, a paradigm shift has occurred in the analysis of medical data \cite{khera2024transforming,ni2024survey}. Techniques such as convolutional neural networks (CNNs) and recurrent neural networks (RNNs) have ushered in significant advancements, demonstrating superior ability in automated ECG classification and the detection of cardiac abnormalities \cite{hammad2018detection,berkaya2018survey}. These deep learning models excel in extracting and learning complex hierarchies of features from raw ECG signals, effectively decoding intricate patterns and temporal dependencies that often elude human experts \cite{liu2021deep}.

Despite these advancements, most current deep learning approaches have been centered on the development of universal models trained on extensive and diverse datasets \cite{ribeiro2020automatic}. Nonetheless, ECG signals exhibit considerable variability across individuals, influenced by distinct factors such as age, gender, body mass index, and genetic predispositions \cite{fratini2015individual}. The presence of noise, artifacts, and various pathologies further complicates ECG signal interpretation \cite{anbalagan2023analysis}. Consequently, these universal models may not perform optimally on an individual basis, potentially leading to inaccuracies and misdiagnoses. 

In response to these limitations, there is an increasing focus on crafting personalized ECG diagnosis strategies tailored to the unique physiological and clinical profile of each patient \cite{li2018patient}. Personalized ECG diagnostics strive to enhance diagnostic precision by customizing models to align closely with the individual's specific ECG characteristics and contextual health information \cite{ebrahimi2020review}. Pioneering work in this area has leveraged advanced deep learning techniques, including transfer learning, domain adaptation, and few-shot learning, showing promising results in fostering personalized diagnostic approaches.

This review conducts a thorough examination of the existing literature on personalized ECG diagnosis leveraging deep learning. It delves into cutting-edge methodologies like fine-tuning, domain adaptation, generative adversarial networks (GANs), diffusion models, and meta-learning. This synthesis not only highlights the capabilities and limitations of these methods but also identifies fertile areas for future inquiry. Our goal is to provide a detailed, critical perspective on how deep learning can revolutionize personalized ECG diagnostics, thereby enhancing the accuracy, efficiency, and patient-centricity of cardiac care.

The structure of this review is as follows: Section II outlines the search methodology employed for literature collection. Section III discusses various deep-learning techniques applied to personalized ECG diagnosis. Section IV explores the challenges and future directions in the field. Section V concludes the review, emphasizing the transformative potential of personalized ECG diagnosis in clinical settings.

\section{Search Methodology}

A rigorous search methodology was employed to ensure a comprehensive and systematic review of the deep-learning techniques applied to personalized ECG diagnosis. The primary databases used for the literature search included PubMed, IEEE Xplore, Web of Science, and Google Scholar, selected for their extensive coverage of biomedical and engineering research, ensuring a broad and inclusive search. The search was conducted over a specified timeframe from January 1, 2020, to May 31, 2024, to capture the most recent advancements in the field. A combination of search terms and keywords was utilized, including “Patient-specific ECG,” “Personalized ECG,” “Domain adaptation ECG,” “Transfer learning ECG,” “GAN ECG,” “Few-shot learning ECG,” and “deep-learning ECG,” among others. Boolean operators (AND, OR) refined the search results, ensuring that relevant studies were not overlooked. Specific inclusion and exclusion criteria were applied to maintain the relevance and quality of the reviewed studies: only peer-reviewed journals and conference articles focused on deep-learning techniques for ECG diagnosis, studies involving human subjects or clinically relevant ECG data, and publications in English were included. Non-peer-reviewed articles, studies focusing on non-human ECG data, and articles not directly related to personalized ECG diagnosis or deep-learning applications were excluded. The initial search yielded a large number of articles, prompting a two-step screening process: the titles and abstracts of the retrieved articles were screened to identify studies meeting the inclusion criteria, followed by a full-text review of the remaining articles to ensure sufficient information on the application of deep-learning techniques to personalized ECG diagnosis. For each selected study, key information was extracted, including study objectives, a description of the deep-learning techniques used, data sources and characteristics, methodological details, and main findings. The extracted data were synthesized to provide a comprehensive overview of the current state of research in personalized ECG diagnosis using deep learning, identifying trends, common methodologies, and gaps in the literature. To ensure reliability and validity, a quality assessment was conducted based on study design, sample size, methodological rigor, and robustness of findings, critically appraising studies to identify potential biases and limitations. 

While the search methodology was designed to be thorough, certain limitations include the exclusion of non-English articles, reliance on specific databases that may have omitted relevant studies, and the fast-evolving nature of deep-learning research, which means recent advancements may not be fully captured. Despite these limitations, the methodology employed provides a robust framework for reviewing the application of deep-learning techniques to personalized ECG diagnosis, offering valuable insights into current practices and future directions in this field.

\begin{figure}[ht]
    \centering
    \includegraphics[width=1.0\textwidth, keepaspectratio]{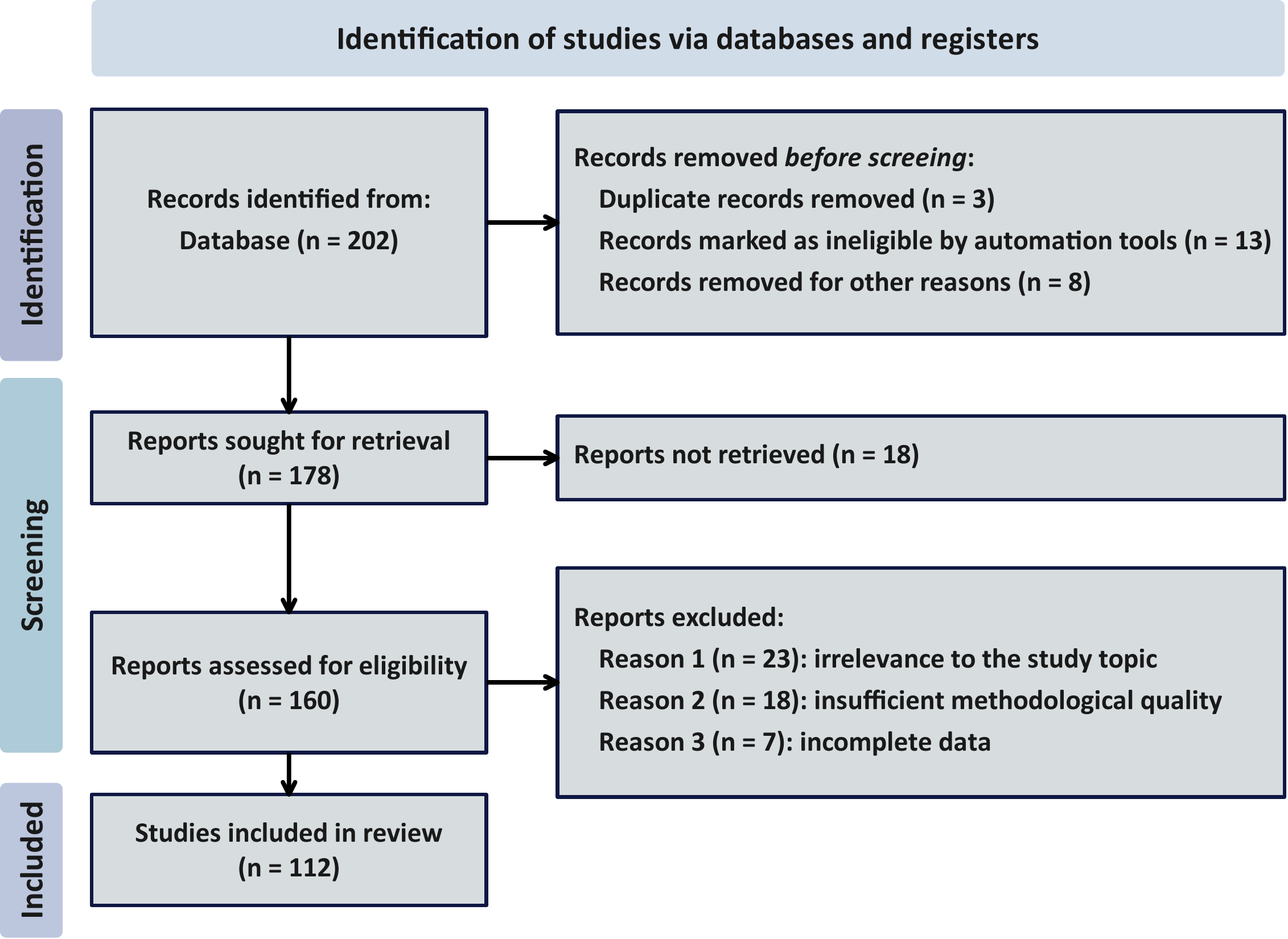}
    \caption{Preferred reporting items for Systematic Reviews and Meta-Analyses (PRISMA) flow diagram showing the number of articles searched and excluded at each stage of the literature search after screening titles, abstracts and full texts.}
    \label{fig:Flowchart}
\end{figure}

\section{Advanced Deep-Learning Techniques for Personalized ECG Diagnosis}

\begin{figure}[ht]
    \centering
    \includegraphics[width=1.0\textwidth, keepaspectratio]{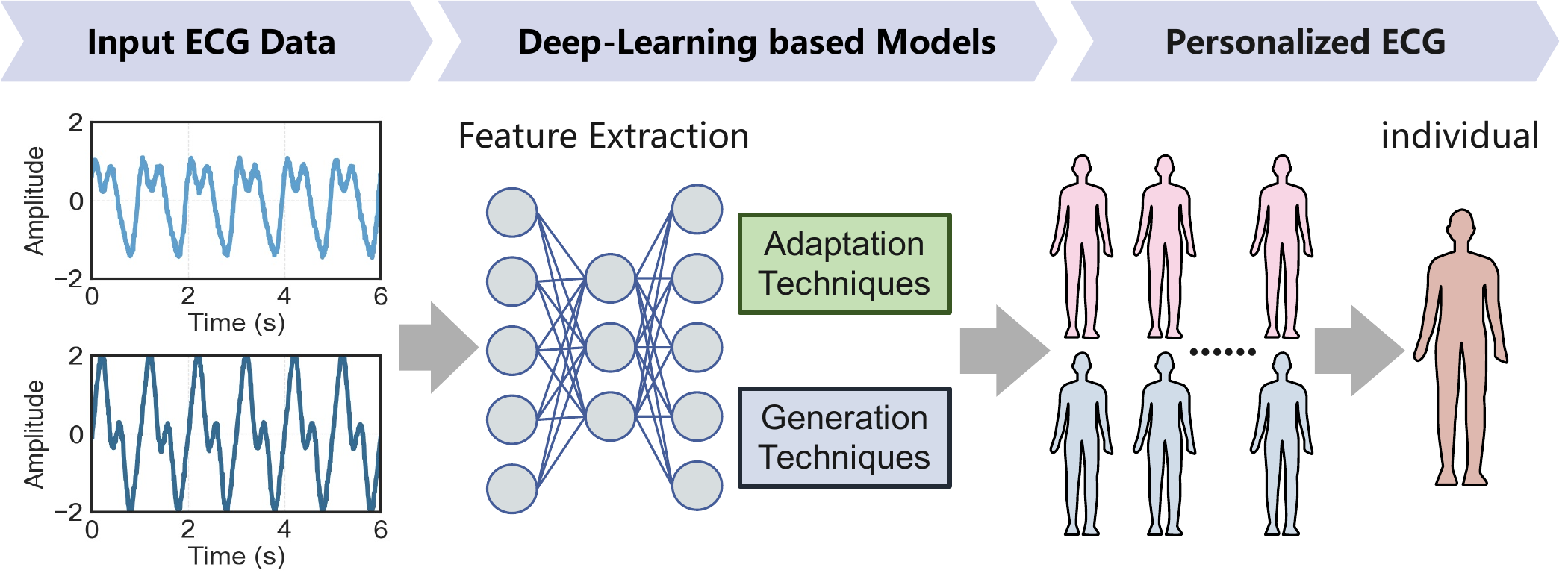}
    \caption{Schematic overview of deep learning-based models for personalized ECG analysis.}
    \label{fig:Overview}
\end{figure}

This review explores the burgeoning field of personalized electrocardiogram (ECG) analysis, which has been revolutionized by the transformative potential of deep learning. As illustrated in Figure \ref{fig:Overview}, the key steps involved in this process include: (1) input ECG data acquisition, (2) application of deep learning models, and (3) implementation of personalized techniques. The paper delves into each of these steps, providing a comprehensive overview of the current state-of-the-art techniques and methodologies. Moreover, it outlines promising avenues for future research that have the potential to further advance this field. By presenting an in-depth exploration of these critical aspects, this review aims to shed light on the latest developments and future directions in personalized ECG analysis driven by deep learning.

Historically, ECG diagnosis has relied heavily on expert interpretation and laborious manual feature extraction. However, the advent of sophisticated deep-learning models is revolutionizing this paradigm. These AI-powered algorithms are increasingly augmenting, and in certain instances, supplanting traditional methods, offering the potential for tailored diagnostic insights and ultimately, improved patient outcomes \cite{siontis2021artificial}. This shift towards automated and personalized analysis promises to enhance diagnostic accuracy, streamline clinical workflows, and facilitate the development of more targeted and effective treatment strategies.

To achieve truly personalized ECG analysis, this review will further explore several cutting-edge techniques. These approaches represent significant advancements in the field, enabling the development of highly customized diagnostic and predictive models that can account for individual variability and potentially unlock new frontiers in personalized cardiovascular medicine.

\subsection{Fine-Tuning}
\subsubsection{Tailoring Pre-trained Models for Personalization}
Fine-tuning is a pivotal transfer learning technique employed to adapt pre-trained neural networks for personalized ECG diagnosis \cite{weimann2021transfer}. This process enhances the model's effectiveness in capturing both general cardiovascular features and patient-specific nuances. Initially, a neural network is trained on a large, diverse dataset, enabling it to learn broad patterns and representations \cite{pal2021cardionet,maray2023transfer}.

\begin{figure*}[ht]
\centering
\includegraphics[width=1\textwidth, keepaspectratio]{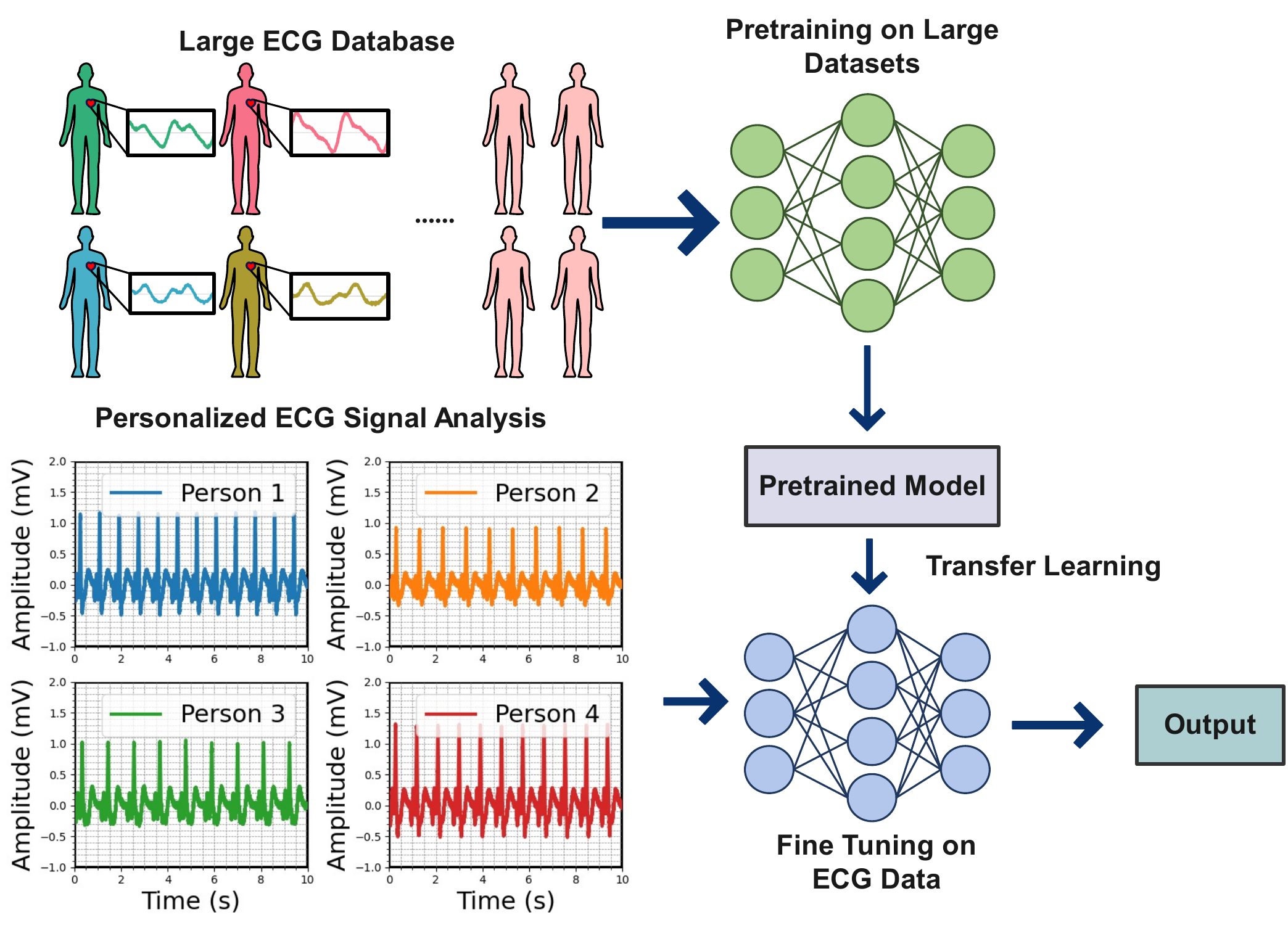}
\caption{Schematic illustration of fine-tuning a pre-trained neural network for personalized ECG diagnosis. The pre-trained network is adapted for patient-specific ECG diagnosis through further training on a smaller dataset, enabling the model to capture unique electrophysiological patterns and improve diagnostic precision.}
\label{fig:FT}
\end{figure*}

Subsequently, as depicted in Fig. \ref{fig:FT}, the pre-trained network undergoes fine-tuning, where it is further trained on a smaller, patient-specific dataset \cite{mohebbanaaz2022new}. This step allows the model to adapt and specialize in the unique electrophysiological patterns exhibited by individual patients. By fine-tuning the pre-trained model, a balance can be striked between retaining general cardiovascular knowledge and honing its ability to detect subtle, personalized variations in ECG signals.

The fine-tuning process involves updating the weights of the pre-trained model based on the patient-specific dataset. This adaptation enables the model to become more sensitive to the unique characteristics of the patient's ECG, addressing inter-patient variability. As a result, the model's diagnostic accuracy and precision are improved, leading to more tailored and effective ECG diagnostics. Fine-tuning allows for a seamless transition from general cardiovascular understanding to personalized healthcare solutions.

By leveraging pre-trained models and fine-tuning them for specific patients, the model's initial broad knowledge is harnessed while its expertise is customized to cater to individual needs. This approach enhances the model's ability to detect anomalies, predict cardiovascular events, and provide personalized recommendations.

\subsubsection{Applications of Fine-Tuning Methods in Personalized ECG Signal Processing}
Fine-tuning methods have become essential in enhancing the performance of personalized ECG signal processing models, enabling them to adapt to individual variations in cardiac rhythms. For instance, Ng et al. \cite{ng2023few} proposed a personalized atrial fibrillation (AF) detector using a Siamese network, which leverages few-shot learning to address the imbalanced dataset problem during fine-tuning. Similarly, Hu et al. \cite{hu2023personalized} explored the impact of different deep-learning model structures and label mapping lengths on personalized transfer learning for single-lead ECG-based sleep apnea detection, demonstrating the superiority of a hybrid transformer model. In another study, Hu et al. \cite{hu2023semi} introduced a semi-supervised learning approach for low-cost personalized obstructive sleep apnea detection, utilizing a convolutional neural network-based auto-encoder to assign pseudo-labels and improve model performance. Hong et al. \cite{hong2021deep} developed a deep-learning model with individualized fine-tuning for dynamic and beat-to-beat blood pressure estimation, showing significant reductions in mean absolute errors. Suhas et al. \cite{suhas2024end} presented an end-to-end personalized cuff-less blood pressure monitoring system using transformers and contrastive loss-based training, achieving state-of-the-art performance even with limited subject-specific data. Liu et al. \cite{liu2024etp} introduced ECG-Text Pre-training (ETP) to learn cross-modal representations linking ECG signals with textual reports, excelling in zero-shot classification tasks. Jia et al. \cite{jia2021learning} proposed a meta-learning algorithm for personalizing a 1D-CNN for ventricular arrhythmias detection on intracardiac EGMs, achieving improved sensitivity and specificity. Additionally, Jia et al. \cite{jia2021device} incorporated prior knowledge into the fine-tuning process to enhance AF detection accuracy, while Ullah et al. \cite{ullah2022automatic} utilized a pre-trained residual network for automatic premature ventricular contraction recognition, demonstrating high accuracy on imbalanced datasets. These studies collectively highlight the effectiveness of fine-tuning methods in adapting deep-learning models to individual patient data, thereby improving the accuracy and reliability of personalized ECG diagnostics.

However, a significant limitation remains in the requirement of manually annotated ECG signals for fine-tuning, particularly for long-term recordings with substantial individual differences. While pre-training AF detection models on public databases and fine-tuning them with a portion of the test data can enhance model generalization, the necessity for manual annotation poses a considerable challenge. This process can be time-consuming and labor-intensive, especially when dealing with extensive and personalized ECG datasets, as highlighted by Ma \cite{ma2023atrial}. Addressing this limitation requires the development of more efficient annotation techniques or the integration of semi-supervised and unsupervised learning methods to reduce the dependency on extensive manual labeling.

\subsection{Domain Adaptation}
\subsubsection{Bridging Data Discrepancies for Better Generalization}
Domain adaptation techniques are crucial for transferring knowledge from a general source domain to a specific target domain, such as from a large, heterogeneous ECG dataset to the personalized ECG data of individual patients. Methods such as adversarial training or discrepancy-based approaches are utilized to align the feature distributions between these domains. By mitigating the discrepancies in data distribution, domain adaptation ensures that the diagnostic model can generalize effectively to the unique ECG patterns of different patients, thereby accommodating inter-patient variability and enhancing the model’s adaptability and accuracy in personalized settings.

\begin{figure}[ht]
    \centering
    \includegraphics[width=0.8\textwidth, keepaspectratio]{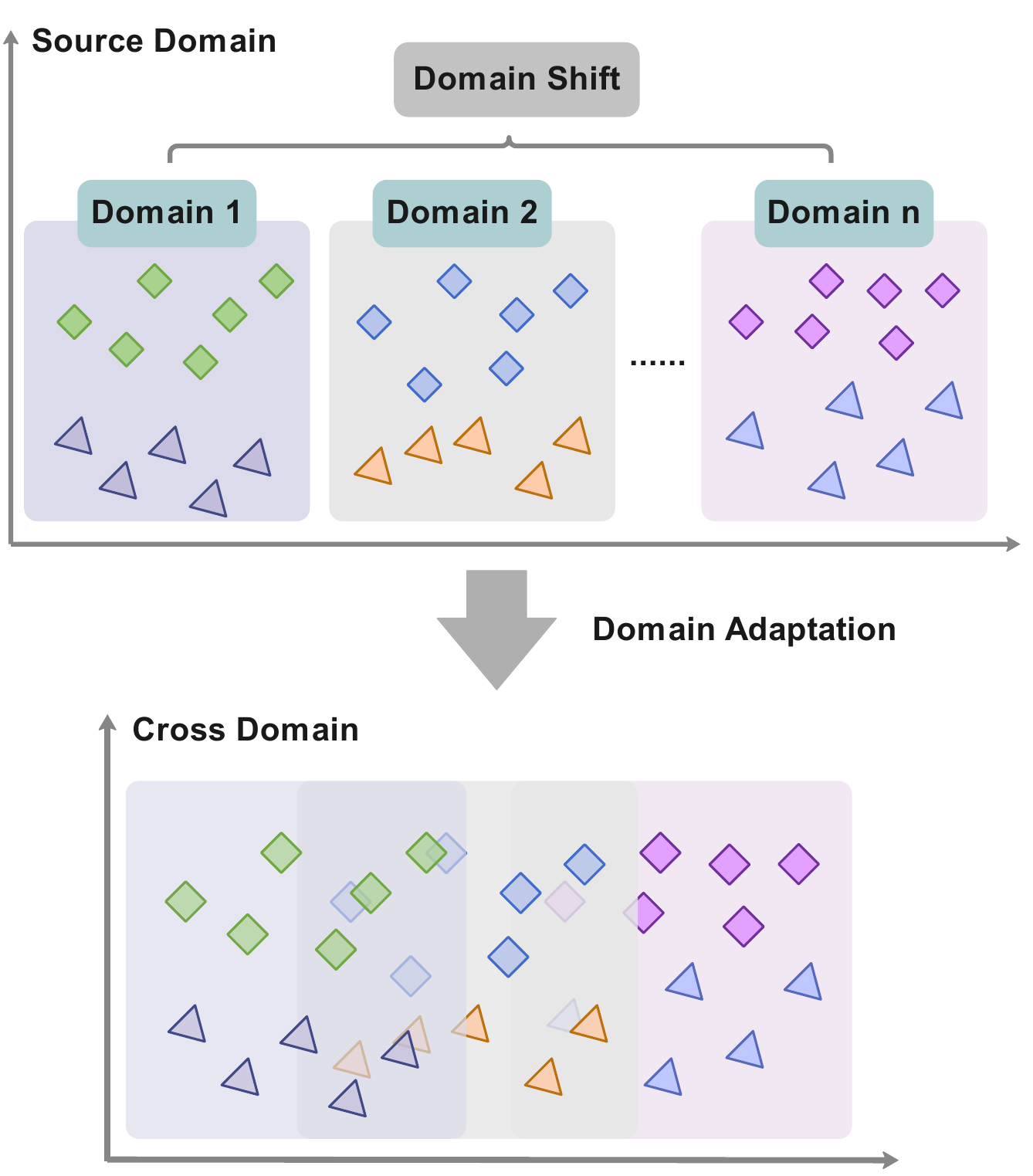}
    \caption{Schematic illustration of domain adaptation for personalized ECG diagnosis. The source domain represents a large, heterogeneous ECG dataset, while the target domain represents the personalized ECG data of an individual patient. Domain adaptation aligns the feature distributions between these domains, enabling the model to generalize effectively to the unique ECG patterns of the target patient.}
    \label{fig:DA}
\end{figure}

Fig. \ref{fig:DA} illustrates the concept of domain adaptation for personalized ECG diagnosis. The source domain represents a large, heterogeneous ECG dataset, while the target domain represents the personalized ECG data of an individual patient. By aligning the feature distributions between these domains, domain adaptation enables the model to generalize effectively to the unique ECG patterns of the target patient, improving its accuracy and adaptability in personalized settings.

\subsubsection{Applications of Domain Adaptation Methods in Personalized ECG Signal Processing}
Domain adaptation methods have been increasingly applied to address the challenges of inter-subject variability and domain shifts in personalized ECG signal processing. He et al. \cite{he2021online} proposed an online cross-subject emotion recognition approach using unsupervised domain adaptation (UDA) to handle both inter-subject and intra-subject discrepancies in ECG signals, demonstrating significant improvements in classification accuracy and robustness in online scenarios. Wang et al. \cite{wang2021inter} introduced a Domain-Adaptative ECG Arrhythmia Classification (DAEAC) model that employs convolutional networks and UDA to improve inter-patient heartbeat classification without requiring expert annotations, leveraging cluster-aligning and cluster-maintaining losses to enhance feature discriminability and structural information. Similarly, Chen et al. \cite{chen2020unsupervised} developed a Multi-path Atrous Convolutional Network (MACN) with cluster-aligning and cluster-separating losses to address performance degradation in inter-patient ECG classification. Yuan and Siyal \cite{yuan2023target} proposed a Target-oriented Augmentation Privacy-protection Domain Adaptation (TAPDA) framework, utilizing source-free domain adaptation (SFDA) to enhance ECG beat classification while preserving patient privacy. Their method addresses class imbalance by employing data augmentation techniques and a class-balance pseudo-labeling strategy, significantly improving classification performance.  Du et al. \cite{du2023diagnosis} proposed a dual-channel network with feature queue technology for atrial fibrillation diagnosis, achieving high precision and recall using unsupervised domain adaptation. Their innovative feature queue technique enables stable and rapid updates of the network, leveraging both labeled and unlabeled data effectively. Feng et al. \cite{feng2022unsupervised} presented an unsupervised semantic-aware adaptive feature fusion network (USAFFN) to reduce domain shifts in arrhythmia detection by aligning semantic distributions between feature spaces. Their multi-perspective adaptive feature fusion (MPAFF) module extracts informative ECG representations from different angles, improving detection performance and generalization across datasets. These studies collectively highlight the effectiveness of domain adaptation methods in enhancing the generalization and robustness of personalized ECG diagnostics, addressing the variability and privacy concerns associated with ECG signal processing.

\subsection{Generative Adversarial Networks (GANs)}
\subsubsection{Synthesizing Realistic Patient-Specific ECG Data}
Generative Adversarial Networks (GANs) have emerged as a powerful paradigm for synthesizing highly personalized ECG data. The GAN architecture comprises two neural networks—a generator and a discriminator—engaged in an adversarial training process. In the context of personalized ECG analysis, the generator synthesizes realistic, patient-specific ECG signals, while the discriminator evaluates their authenticity, ensuring that the generated signals faithfully capture individual cardiac characteristics.

\begin{figure}[ht]
    \centering
    \includegraphics[width=0.8\textwidth, keepaspectratio]{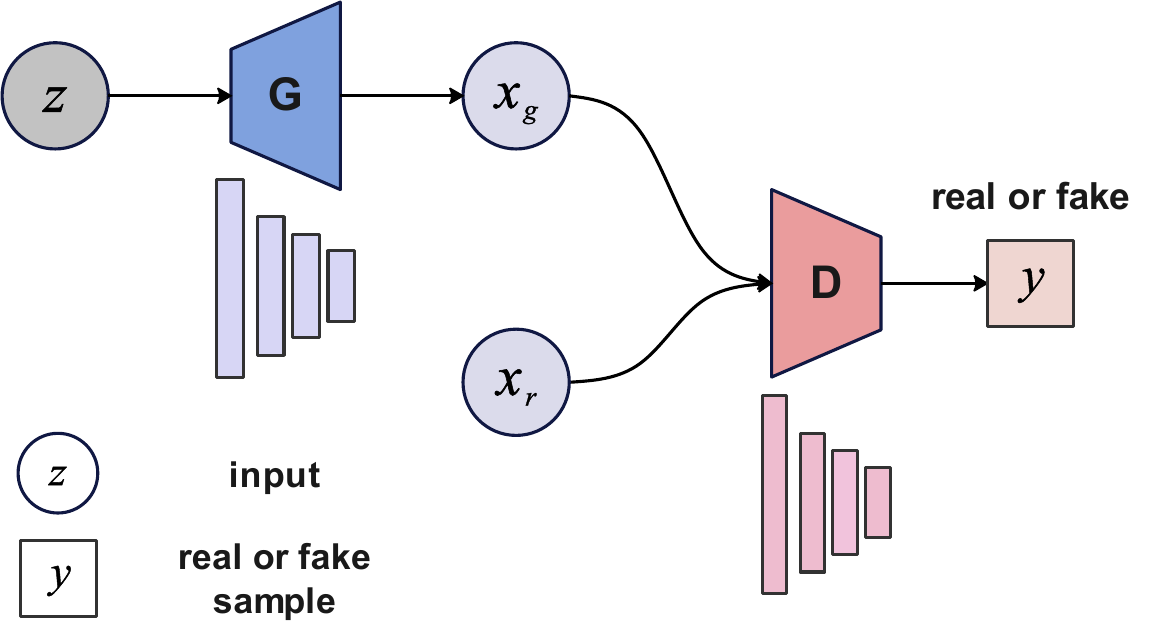}
    \caption{Schematic illustration of a Generative Adversarial Network (GAN). A Generator (G) synthesizes data aiming to mimic the true data distribution, and a Discriminator (D) evaluates the authenticity of the generated data against real samples. The Generator receives a random noise vector (z) as input and produces an output that attempts to pass as a real sample. The Discriminator, trained to distinguish between real data and generated samples, provides feedback to the Generator to improve its synthesis process. This adversarial training framework pushes the Generator to produce increasingly realistic data over time.}
    \label{fig:GAN}
\end{figure}

Fig. \ref{fig:GAN} presents a schematic illustration of a GAN architecture tailored for personalized ECG signal generation. The generator network, typically comprising convolutional neural networks (CNNs) or recurrent neural networks (RNNs), learns to produce synthetic ECG signals that meticulously mimic the unique characteristics of individual patient data. Concurrently, the discriminator network distinguishes between real patient-specific ECG signals and those generated by the model. This adversarial interplay drives the generator to create increasingly realistic and personalized ECG data, while the discriminator becomes more adept at detecting subtle, patient-specific differences.

Recent advancements in GAN-based approaches have shown promising results for personalized ECG synthesis and analysis. Golany et al. \cite{golany2019pgans} introduced PGANs (Personalized Generative Adversarial Networks), which learn to generate patient-specific ECG signals by conditioning the generator on individual patient characteristics. This approach significantly improved the quality and personalization of synthetic ECG data, enhancing diagnostic accuracy for patient-specific arrhythmia detection. Similarly, Delaney et al. \cite{delaney2019synthesis} proposed a GAN-based method for synthesizing realistic, patient-specific 12-lead ECG signals, successfully capturing intricate correlations between different ECG leads while preserving individual patient characteristics.

In the domain of personalized ECG diagnosis, GANs have demonstrated remarkable capability in generating high-fidelity synthetic ECG data that closely mimics a patient's unique cardiac patterns. This synthetic data serves as a valuable augmentation to limited patient-specific datasets, significantly enriching the training set and enhancing the robustness and accuracy of personalized diagnostic models. By meticulously capturing the nuances of individual patients' ECGs, GANs effectively address inter-patient variability and contribute to more personalized and precise diagnostic outcomes.

The application of GANs in personalized ECG analysis extends beyond data synthesis. For instance, Shaker et al. \cite{shaker2020generalization} utilized GANs to address the challenge of generalization in ECG classification across different patients, improving the performance of ECG classification models on unseen patients and enhancing the personalization of diagnostic systems.

In synopsis, GAN-generated, high-fidelity, patient-tailored synthetic ECGs critically empower personalized diagnostic model development, heralding transformative potential for individualized cardiac care and outcome optimization, with anticipated breakthroughs forthcoming as this synergistic research frontier continues to evolve.

\subsubsection{Leveraging GAN-Based Models for Enhanced Personalized ECG Signal Processing}

GANs have been extensively utilized in personalized ECG signal processing to address challenges such as patient-specific data scarcity, individual privacy concerns, and the need for high-quality synthetic data that captures unique cardiac characteristics. Chen et al. \cite{chen2022me} introduced ME-GAN, a disease-aware GAN for multi-view ECG synthesis conditioned on individual heart diseases, which significantly improves the representation of patient-specific ECG signals by precisely injecting personalized disease information into suitable locations. This approach enhances the model's ability to capture individual variations in cardiac pathology.

Golany et al. \cite{golany2021ecg} proposed ECG ODE-GAN, which learns the dynamics of personalized ECG signals through ordinary differential equations, enhancing ECG heartbeat classification for individual patients. This method adapts to the unique temporal characteristics of each patient's cardiac signals, improving diagnostic accuracy \cite{golany2019pgans}. Neifar et al. \cite{neifar2024leveraging} leveraged statistical shape priors in GAN-based ECG synthesis to generate realistic, patient-specific ECG signals, effectively addressing class imbalance in individual training datasets.

Sarkar and Etemad \cite{sarkar2021cardiogan} developed CardioGAN, which synthesizes personalized ECG from PPG signals using an attention-based generator and dual discriminators. This approach significantly improves heart rate measurement accuracy for individual patients by capturing patient-specific correlations between PPG and ECG signals. Kim and Pan \cite{kim2021study} utilized an auxiliary classifier GAN to generate synthetic ECG signals for user recognition, demonstrating high recognition performance even with inconsistent data sizes across individuals.

Golany et al. \cite{golany2020simgans} introduced SimGANs, integrating a biological simulator into the GAN framework to generate biologically plausible, patient-specific ECG training examples. This innovative approach improves heartbeat classification by incorporating individual physiological constraints. Kang et al. \cite{kang2023gan} demonstrated a multi-task GAN model for ECG authentication and arrhythmia detection, effectively hiding sensitive health information while preserving individual cardiac characteristics.

Wang et al. \cite{wang2023hierarchical} presented a hierarchical deep-learning framework with GAN for automatic cardiac diagnosis from ECG signals, addressing data-lacking and imbalanced issues in personalized datasets. Byeon and Kwak \cite{byeon2023semi} leveraged semi-supervised domain adaptation with CycleGAN to account for electrode placement discrepancies in individual identification from ECG signals, enhancing the robustness of personalized ECG analysis.

Rafi and Ko \cite{rafi2023sf} introduced SF-ECG, a source-free domain adaptation approach using GANs to synthesize patient-specific ECG data and improve arrhythmia classification. This method adapts to individual variations in ECG morphology, enhancing the personalization of arrhythmia detection algorithms. Vo et al. \cite{vo2021p2e} developed P2E-WGAN, a conditional Wasserstein GAN for synthesizing personalized ECG waveforms from PPG, demonstrating effectiveness in augmenting patient-specific training data.

Joo et al. \cite{joo2023twelve} developed a GAN to reconstruct 12-lead ECG signals from single-lead inputs, enhancing the detection of complex cardiovascular diseases by preserving individual patient characteristics across leads. Kuo et al. \cite{kuo2022towards} applied a self-attention GAN for personalized sleep scoring, significantly improving classification accuracy by capturing individual sleep patterns in ECG signals.

These studies collectively highlight the effectiveness of GAN-based models in generating high-quality, patient-specific synthetic ECG data, improving personalized classification performance, and addressing privacy concerns in individualized ECG signal processing. The ability of GANs to capture and replicate unique cardiac characteristics of individual patients represents a significant advancement in personalized ECG analysis, paving the way for more accurate and tailored diagnostic and monitoring solutions in cardiology.
\subsection{Diffusion Models}
\subsubsection{Generating High-Fidelity Personalized ECG Signals}
Diffusion models, a class of generative models that reverse a stochastic diffusion process, have emerged as a powerful tool for generating high-fidelity ECG signals \cite{ho2020denoising}. In the realm of personalized ECG diagnosis, these models excel at learning the underlying distribution of patient-specific heartbeat patterns. By iteratively refining noisy ECG signals, diffusion models produce realistic, individualized data that accurately capture the unique cardiac characteristics of each patient. This capability is particularly advantageous in scenarios with limited or noisy patient data, as diffusion models can augment training datasets with high-quality synthetic signals that preserve patient-specific features \cite{ouyang2024transfer,debnath2024impact}. Consequently, these models effectively address inter-patient variability and significantly enhance the personalization of diagnostic algorithms.

\begin{figure}[ht]
    \centering
    \includegraphics[width=0.8\textwidth, keepaspectratio]{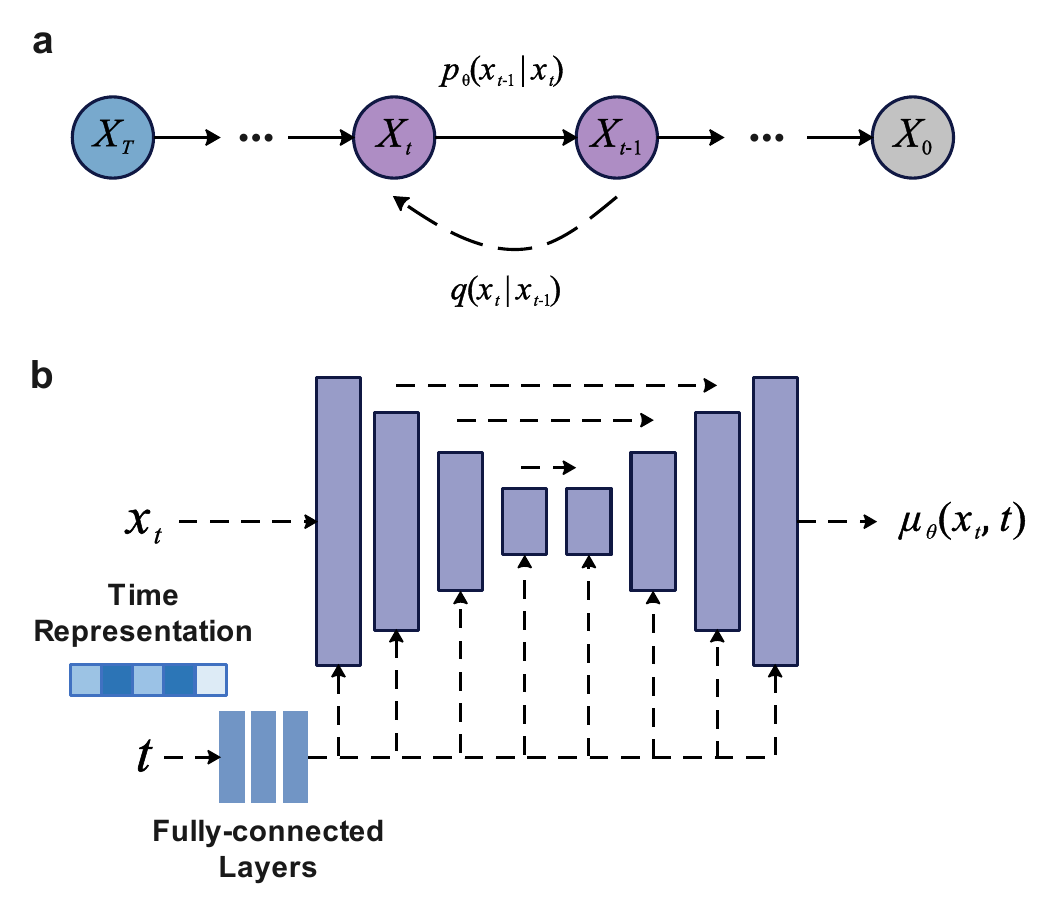}
    \caption{Illustration of the diffusion process in Diffusion Models. (a) The forward diffusion process, where $x_0$ transitions through a Markov chain to $x_t$, depicts the stochastic process of transitioning from one state to another in the state space, with time information synchronously embedded at each step. (b) A common architecture of a diffusion model utilizes a UNet structure, residual blocks, and skip connections, as well as embedding time information at various layers to facilitate the reverse diffusion process. This architecture efficiently reconstructs images from the noise distribution back to the data distribution by synchronously incorporating temporal embeddings.}
    \label{fig:Diffusion}
\end{figure}

Fig. \ref{fig:Diffusion} illustrates the architecture and functionality of a diffusion model tailored for ECG signal processing. The model incorporates a sophisticated combination of UNet architecture, ResNet blocks, skip connections, and self-attention layers to process noisy ECG data and generate high-fidelity, patient-specific signals. The input ECG data, containing noise and artifacts, undergoes a series of denoising steps, where the model progressively refines the signal while preserving individual patient characteristics. Through the utilization of residual blocks and skip connections, the model effectively captures both local and global dependencies within the ECG data, ensuring that important temporal features and patient-specific patterns are preserved.

\subsubsection{Applications of Diffusion Models in Personalized ECG Signal Processing}
Recent studies have demonstrated the significant potential of diffusion models in advancing personalized ECG diagnosis through various tasks, including patient-specific synthetic ECG generation, individualized noise and baseline wander removal, and cross-modal signal-to-signal translation tailored to individual patients. Adib \emph{et al.} \cite{adib2023synthetic} introduced an innovative pipeline utilizing Improved Denoising Diffusion Probabilistic Models (DDPM) for generating synthetic ECG signals by embedding 1D ECG time series data into 2D space. This approach enables the generation of highly personalized ECG signals, thereby enhancing the training dataset and supporting the development of individualized diagnostic systems.

Lopez Alcaraz and Strodthoff \cite{alcaraz2023diffusion} proposed SSSD-ECG, a diffusion model underpinned by structured state-space models for generating conditional 12-lead ECGs customized to individual patient profiles. This model demonstrates superior performance in capturing patient-specific ECG characteristics compared to traditional generative adversarial networks (GANs). Zama and Schwenker \cite{zama2023ecg} developed a diffusion-based model coupled with a State Space Augmented Transformer for synthesizing conditional 12-lead ECGs based on the PTB-XL dataset \cite{wagner2020ptb}. Their approach significantly enhances personalization by learning distinct patient-specific signal characteristics, enabling the generation of highly individualized ECG signals.

Shome \emph{et al.} \cite{shome2024region} introduced the Region-Disentangled Diffusion Model (RDDM) for translating Photoplethysmography (PPG) signals to ECG. This innovative approach enables cross-modal signal translation that can improve patient-specific monitoring by accurately capturing individual cardiovascular characteristics across different modalities. Li \emph{et al.} \cite{li2023descod} proposed the Deep Score-Based Diffusion Model for ECG Baseline Wander and Noise Removal (DeScoD-ECG), offering personalized noise reduction by adapting to individual signal characteristics. This model achieved at least a 20\% overall improvement compared to the best baseline method, providing cleaner and more accurate patient-specific signals for enhanced diagnostic accuracy.

A key advantage of diffusion models in personalized ECG analysis lies in their ability to capture the complex temporal dynamics and individual variations of ECG signals, enabling robust personalization. By incorporating advanced techniques such as embedding 1D ECG time series data into 2D space \cite{adib2023synthetic}, utilizing structured state-space models \cite{alcaraz2023diffusion}, and employing region-disentangled diffusion processes \cite{shome2024region}, these models can generate high-quality synthetic ECG signals that closely resemble real ECG recordings while accurately reflecting individual patient differences. This capability is crucial for developing personalized diagnostic and monitoring systems that can adapt to the unique cardiovascular profiles of individual patients.
\subsection{Meta-Learning}
\subsubsection{Rapid Adaptation with Minimal Data}
Few-shot learning and meta-learning techniques are designed to enable models to learn new tasks rapidly with minimal data, which is critical for personalized ECG diagnosis. These approaches allow diagnostic models to adapt to a new patient's ECG data using only a few examples. Meta-learning frameworks, such as Model-Agnostic Meta-Learning (MAML), train models on a variety of tasks, equipping them with the ability to quickly adjust to new, unseen tasks. This rapid adaptability is essential for personalized diagnosis, as it enables the model to accurately interpret the unique ECG patterns of individual patients, even with limited training data. By addressing inter-patient variability, few-shot learning and meta-learning significantly enhance the model's ability to provide personalized and precise diagnostic insights.

\begin{figure}[ht]
\centering
\includegraphics[width=0.52\textwidth, keepaspectratio]{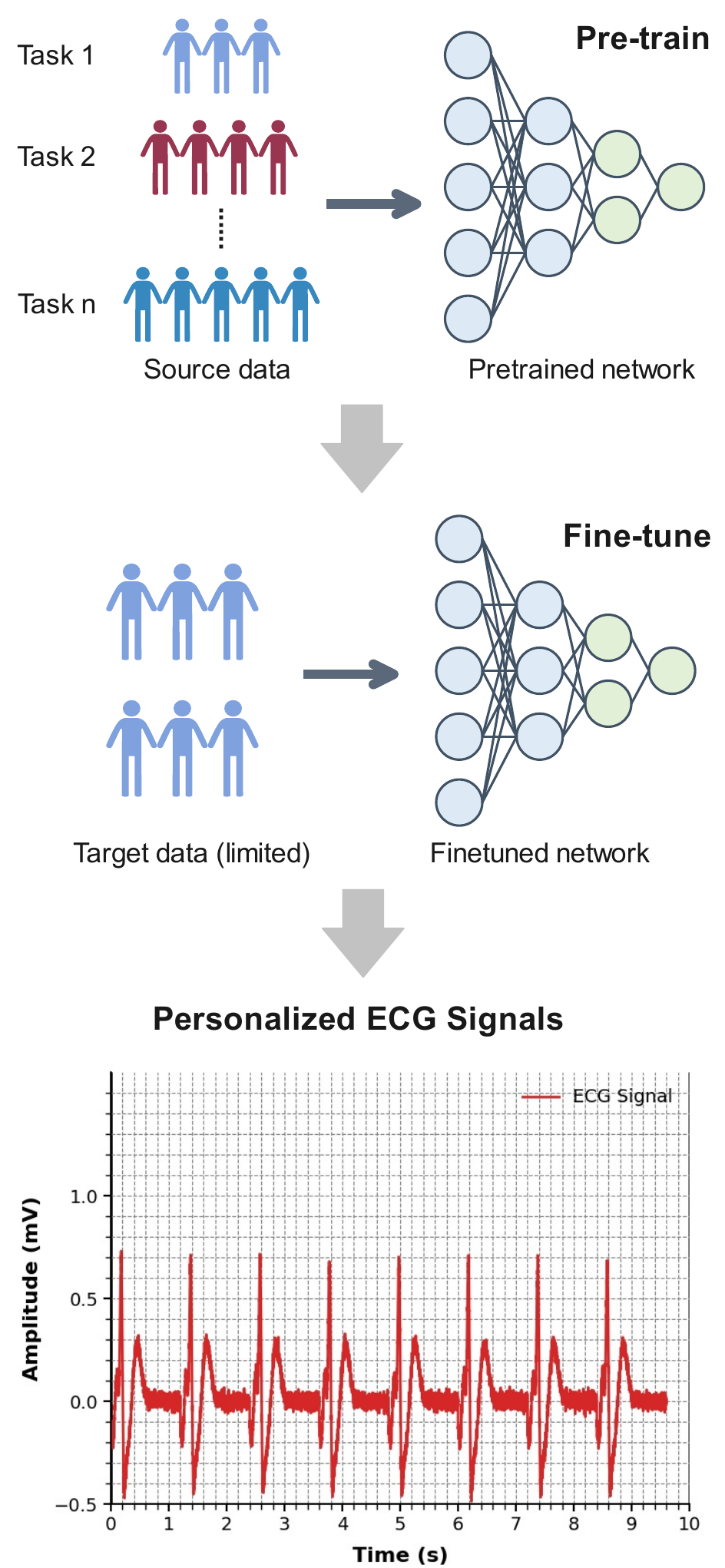}
\caption{Schematic illustration of meta-learning for personalized ECG diagnosis. The model is trained on a variety of tasks, each representing a different patient's ECG data, and can be fine-tuned for a new patient's data using only a few examples.}
\label{fig:Meta}
\end{figure}

Fig. \ref{fig:Meta} illustrates the concept of meta-learning for personalized ECG diagnosis. The model is trained on a variety of tasks, each representing a different patient's ECG data. This training enables the model to learn a generalizable representation of ECG signals, which can be fine-tuned for a new patient's data using only a few examples. This rapid adaptation enables the model to accurately diagnose the patient's condition, even with minimal training data.

\subsubsection{Applications of Meta-Learning-based Model in Personalized ECG Signal Processing}
Meta-learning approaches have shown significant promise in enhancing personalized ECG signal processing by addressing individual variability and the scarcity of labeled data. Akbari et al. \cite{akbari2021meta} introduced a meta-learning framework for translating bio-impedance (Bio-Z) signals to ECG, which adapts to new users with minimal samples, reducing the percentage root mean square difference (PRD) substantially compared to traditional methods. Jia et al. \cite{jia2022personalized} proposed a personalized neural network for patient-specific health monitoring in IoT, leveraging a meta-learning method to quickly adapt a well-generalized model to new patient data, achieving notable improvements in VF detection, AF detection, and human activity recognition. Liu et al. \cite{liu2023diagnosis} developed a Meta Siamese Network (MSN) for few-shot ECG diagnosis, effectively improving the robustness of arrhythmia detection with limited records by employing a metric-based meta-learning framework. Meqdad et al. \cite{meqdad2022meta} proposed an interpretable meta-structural learning algorithm for arrhythmia detection across multiple sessions, achieving high accuracy and competitive performance by combining CNNs and genetic programming. Musa et al. \cite{musa2023systematic} conducted a comprehensive systematic review and meta-data analysis on the applications of deep-learning in ECG signal processing, highlighting the various domains and challenges, and providing insights into potential research opportunities. Sun et al. \cite{sun2023federated} proposed a federated learning and blockchain framework for physiological signal classification, integrating meta-learning to address dynamic distribution changes and privacy concerns, demonstrating superior performance and privacy protection.

Liu et al. \cite{liu2021few} introduced a meta-transfer-based few-shot learning method for arrhythmia detection using ECG data from wearable devices, which effectively handles the data scarcity issue by converting ECG signals into spectrograms and leveraging a pre-trained feature extractor. Rjoob et al. \cite{rjoob2022machine} reviewed the evolution of machine learning applications in ECG over two decades, noting the increasing prominence of deep-learning techniques and their superior performance in arrhythmia detection.

Sun et al. \cite{sun2023few} addressed the challenge of classifying medical time series with few-shot class-incremental learning by proposing the Meta self-attention Prototype Incrementer (MAPIC), which significantly outperforms state-of-the-art methods in handling new classes without forgetting old ones. Essa and Xie \cite{essa2021ensemble} developed an ensemble of deep-learning-based multi-model systems for ECG heartbeat classification, combining CNN-LSTM and RRHOS-LSTM models to achieve high accuracy and robustness.

\subsection{Comparative Analysis}
Table \ref{tab:summary} presents a comprehensive overview of recent advancements in deep-learning techniques for personalized ECG diagnosis. This table highlights the significant performance improvements achieved through personalization, demonstrating the effectiveness of these techniques in capturing individual-specific cardiac patterns and enhancing diagnostic accuracy.

\begin{sidewaystable}
\caption{Overview of Recent Advances in Deep-Learning Techniques for Personalized ECG Analysis} \label{tab:summary}
\begin{tabular*}{\textheight}{@{\extracolsep\fill}p{1.5cm}p{1.5cm}p{2cm}p{2cm}p{1.5cm}p{1cm}p{2cm}p{2cm}}
\toprule%
\textbf{Study} & \textbf{Technique} & \textbf{Task} & \textbf{Dataset} & \textbf{Sample Size} & \textbf{Leads} & \textbf{Before Personalization} & \textbf{After Personalization} \\
\midrule
Ng et al. 2023 \cite{ng2023few} & Fine-Tuning & Atrial Fibrillation Detection & Private Dataset & 1,000 & 1 & F1 Score: 95\% & F1 Score: 97\% \\ \midrule
Hu et al. 2023 \cite{hu2023personalized} & Fine-Tuning & Sleep Apnea Detection & MIT-BIH Polysomnographic & 18 & 1 & Accuracy: 0.8412 & Accuracy: 0.8537 \\ \midrule
Hong et al. 2021 \cite{hong2021deep} & Fine-Tuning & Blood Pressure Estimation & Private Dataset & 85 & 1 & MAE: 13.43 mmHg & MAE: 9.49 mmHg \\ \midrule
Suhas et al. 2024 \cite{suhas2024end} & Fine-Tuning & Blood Pressure Monitoring & Private Dataset & 510 & 1 & MAE: 1.5 mmHg & MAE: 1.08 mmHg \\ \midrule
He et al. 2021 \cite{he2021online} & Domain Adaptation & Emotion Recognition & Private Dataset & 32 & 1 & Baseline Performance & ~12\% improvement \\ \midrule
Wang et al. 2021 \cite{wang2021inter} & Domain Adaptation & Arrhythmia Classification & MIT-BIH Arrhythmia & 47 & 2 & Accuracy: 97.0\% & Accuracy: 97.59\% \\ \midrule
Chen et al. 2020 \cite{chen2020unsupervised} & Domain Adaptation & Arrhythmia Classification & MIT-BIH Arrhythmia & 47 & 2 & VEB F1: 88\%, SVEB F1: 70\% & VEB F1: 95\%, SVEB F1: 88\% \\ \midrule
Chen et al. 2022 \cite{chen2022me} & GANs & ECG Synthesis & Private Dataset & 200 & 12 & S: (0.4, 0.39), V: (0.85, 0.87), F: (0.2, 0.07) & S: (0.4, 0.5), V: (0.85, 0.87), F: (0.2, 0.1) \\ \midrule
Golany et al. 2021 \cite{golany2021ecg} & GANs & ECG Heartbeat Classification & Private Dataset & 168 & 2 & S: (0.40/0.70), F: (0.20/0.25), V: (0.90/0.86) & S: (0.40/0.83), F: (0.20/0.45), V: (0.90/0.90) \\
\botrule
\end{tabular*}
\end{sidewaystable}

\begin{sidewaystable}
\begin{tabular*}{\textheight}{@{\extracolsep\fill}p{1.5cm}p{1.5cm}p{2cm}p{2cm}p{1.5cm}p{1cm}p{2cm}p{2cm}}
\multicolumn{8}{c}{\bf TABLE 1 - Continued from previous page}\\
\toprule%
\textbf{Study} & \textbf{Technique} & \textbf{Task} & \textbf{Dataset} & \textbf{Sample Size} & \textbf{Leads} & \textbf{Before Personalization} & \textbf{After Personalization} \\
\midrule
Neifar et al. 2024 \cite{neifar2024leveraging} & GANs & ECG Synthesis & Private Dataset & 250 & 12 & RMSE: N(1.86e-3, 2.45e-2), V(2.05e-3, 2.56e-2), F(2.21e-3, 2.78e-2) & RMSE: N(1.66e-3, 1.84e-2), V(1.95e-3, 2.12e-2), F(2.05e-3, 2.35e-2) \\ \midrule
Sarkar and Etemad 2021 \cite{sarkar2021cardiogan} & GANs & ECG from PPG Synthesis & Private Dataset & 42 & 1 & Heart Rate Error: 9.74 bpm & Heart Rate Error: 2.89 bpm \\ \midrule
Adib et al. 2023 \cite{adib2023synthetic} & Diffusion Models & ECG Generation & MIT-BIH Arrhythmia & 47 & 2 & Average Precision: 0.76 & Average Precision: 0.96 \\ \midrule
Lopez Alcaraz et al. 2023 \cite{alcaraz2023diffusion} & Diffusion Models & Conditional ECG Generation & PTB-XL & 21,837 & 12 & AUROC: 0.5968 & AUROC: 0.8402 \\ \midrule
Zama and Schwenker 2023 \cite{zama2023ecg} & Diffusion Models & Conditional ECG Synthesis & PTB-XL & 21,837 & 12 & Accuracy: 95.01\% & Accuracy: 95.84\% \\ \midrule
Shome et al. 2024 \cite{shome2024region} & Diffusion Models & PPG to ECG Translation & Multiple Datasets & >10,000 & 1 & MAE-DALIA: Baseline, MAE-WESAD: Baseline, F1: Baseline, Accuracy: Baseline & MAE-DALIA: -1.53 bpm, MAE-WESAD: -8.10 bpm, F1: +16\%, Accuracy: +4-15\% \\ \midrule
Li et al. 2023 \cite{li2023descod} & Diffusion Models & Noise Removal & QT Database, MIT-BIH & 155 & 2-12 & Cosine Similarity: 0.897 & Cosine Similarity: 0.926 \\ \midrule
Jia et al. 2022 \cite{jia2022personalized} & Meta-Learning & Health Monitoring & Private Dataset & 250 & 1 & VF Accuracy: Baseline, AF Accuracy: Baseline & VF Accuracy: +8.2\%, AF Accuracy: +2.5\% \\ \midrule
Liu et al. 2023 \cite{liu2023diagnosis} & Meta-Learning & Arrhythmia Detection & MIT-BIH Arrhythmia & 47 & 2 & Accuracy: 96.96\% & Accuracy: 99.34\% \\
\botrule
\end{tabular*}
\end{sidewaystable}

\begin{sidewaystable}
\begin{tabular*}{\textheight}{@{\extracolsep\fill}p{1.5cm}p{1.5cm}p{2cm}p{2cm}p{1.5cm}p{1cm}p{2cm}p{2cm}}
\multicolumn{8}{c}{\bf TABLE 1 - Continued from previous page}\\
\toprule%
\textbf{Study} & \textbf{Technique} & \textbf{Task} & \textbf{Dataset} & \textbf{Sample Size} & \textbf{Leads} & \textbf{Before Personalization} & \textbf{After Personalization} \\
\midrule
Meqdad et al. 2022 \cite{meqdad2022meta} & Meta-Learning & Arrhythmia Detection & Private Dataset & 100 & 1 & Accuracy: 97.15\% & Accuracy: 98.29\% \\ \midrule
Sun et al. 2023 \cite{sun2023few} & Meta-Learning & Time Series Classification & Private Dataset & 1,000 & 1 & Accuracy: Baseline & Accuracy: +27.99\%, +18.4\%, +3.95\% (3 datasets) \\ \midrule
Essa and Xie 2021 \cite{essa2021ensemble} & Meta-Learning & Heartbeat Classification & MIT-BIH Arrhythmia & 47 & 2 & F1 Score: Baseline & F1 Score: +3\%, Accuracy: 95.81\% \\
\botrule
\end{tabular*}
\end{sidewaystable}

This analysis reveals a clear trend towards personalized ECG diagnosis, driven by the success of deep-learning techniques. These techniques can be broadly categorized into two main approaches: Adaptation and Generation, with Meta-learning offering a unique bridge between the two.

Adaptation techniques aim to tailor pre-trained models to individual patient data. This approach utilizes real patient data and leverages transfer learning to quickly learn patient-specific characteristics. Examples include fine-tuning and domain adaptation. Fine-tuning directly adapts a pre-trained model to a target patient's ECG data, enhancing its ability to capture subtle individual variations. Domain adaptation tackles the challenge of transferring knowledge from a general source domain to a specific target domain by aligning feature distributions between heterogeneous datasets and patient-specific data. While effective, both approaches often require a significant amount of labeled patient data, which remains a significant challenge in personalized ECG diagnosis.

Generation techniques address the issue of limited patient data by generating synthetic ECG signals. These techniques leverage generative models to create realistic ECG recordings that capture individual variations. GANs and diffusion models are prominent examples. GANs learn to generate high-fidelity synthetic ECG data by capturing complex patterns from real data. Diffusion models excel in generating high-quality synthetic ECG signals and removing noise from recordings, capturing intricate temporal dynamics. These techniques offer a promising avenue for overcoming data scarcity and enabling more personalized ECG analysis.

Meta-learning stands as a powerful technique for rapid adaptation to new tasks with minimal data, making it an essential tool for personalized ECG diagnosis. This approach trains models on a variety of tasks, equipping them with the ability to quickly adjust to new, unseen tasks. This rapid adaptability is crucial for accurately interpreting the unique ECG patterns of individual patients, even with limited training data. By bridging the gap between adaptation and generation, meta-learning offers a unique path to personalized ECG diagnosis, especially when data is scarce.

In conclusion, the integration of advanced deep-learning techniques for personalized ECG diagnosis holds transformative potential. These techniques not only improve diagnostic accuracy and reliability but also address the inherent inter-individual variability in ECG signals. Future research should focus on developing efficient annotation techniques, enhancing model interpretability, and ensuring the clinical validity of synthetic data. By addressing these challenges, the field can move closer to realizing the full potential of personalized ECG diagnosis in clinical practice.

\section{Challenges}
This section explores the multifaceted challenges shaping the landscape of personalized ECG analysis using deep learning. While this domain harbors immense potential to revolutionize cardiac diagnostics and patient care, it is confronted with a series of technical, ethical, and practical hurdles that must be addressed to fully realize its transformative potential.

\subsection{Real-time Learning and Adaptation}
A critical limitation of current models is their static nature, which relies on pre-existing datasets. True clinical utility, however, requires capabilities for real-time processing and learning from streaming ECG data. Future research should focus on developing dynamic algorithms that not only adapt and update in real-time as new data accrues but also do so with minimal computational overhead. Techniques such as online learning and incremental learning are promising but require significant advancements to handle the high-dimensional, continuous influx of ECG data efficiently. Edge computing could play a crucial role here, facilitating rapid data processing close to the data source (e.g., wearable devices), thus minimizing latency issues \cite{hoi2021online,bonomi2022sharing,ngu2021iot}. Furthermore, implementing federated learning could allow for decentralized model updating across numerous devices while safeguarding patient privacy, offering a balanced approach to collaborative healthcare innovation.

\subsection{Addressing Intra-patient Variability}
Personalized ECG systems must account for the substantial variability in a patient's ECG signals, influenced by daily physiological changes and external factors like activity levels and stress. To capture this intra-patient variability effectively, it is essential to develop robust data collection strategies that gather comprehensive ECG data under varying conditions. Advanced machine learning strategies, such as multi-task learning, contextual modeling and cross-modal learning, should be explored to integrate and leverage this variability. These models could incorporate additional data streams (e.g., from wearable sensors) to enrich the ECG data, providing a more holistic view of the patient's health status at any given time \cite{ni2022cross,ni2022progressive,mehari2023towards,patro2024mamba,ni2024adaptive}.

\subsection{Mitigating Negative Transfer Learning}
Transfer learning is pivotal in personalizing models with limited data but carries the risk of negative transfer, where irrelevant knowledge from the source model may impair learning in the target model. This necessitates the development of sophisticated mechanisms to assess and ensure the relevance and benefit of the transferred features. Adaptive strategies that can discern and extract beneficial knowledge specific to each patient's data profile are required. Incorporating meta-learning could refine these processes by enabling systems to learn how to learn more effectively from limited patient data, ensuring that transfer learning enhances rather than detracts from model performance \cite{hong2020opportunities}.

\subsection{Overcoming Data Scarcity}
The scarcity of patient-specific ECG data poses a significant barrier to effective personalization, particularly for rare cardiac conditions or newly onboarded patients. Innovative solutions such as few-shot learning and sophisticated data augmentation techniques designed for temporal data can help mitigate these issues. Furthermore, the use of synthetic data generation through advanced GANs offers a dual opportunity: augmenting training datasets and preserving privacy. However, this approach requires careful validation to ensure that synthetic data maintain high fidelity to real physiological signals and do not introduce biases or artifacts that could mislead diagnostic models \cite{thambawita2021deepfake, mason2024ai}.

\subsection{Ethical and Regulatory Considerations}
As the boundaries of personalized ECG analysis are pushed further, ethical and regulatory challenges become increasingly pertinent. Ensuring patient consent, data security, and privacy in the age of AI-driven diagnostics is paramount. The development of these technologies must also consider potential biases in training data that could lead to disparities in healthcare outcomes. Regulatory frameworks will need to evolve to address these issues, ensuring that innovations in personalized ECG analysis are both safe and beneficial across diverse patient populations.

Addressing these challenges necessitates a concerted effort from multidisciplinary teams, encompassing data scientists and cardiologists alongside ethicists and policymakers. By confronting these challenges directly, the full potential of personalized ECG analysis can be realized, transforming cardiac care with more accurate, timely, and individualized diagnostics.

\section{Conclusion and Future Perspective}

Despite the numerous challenges that lie ahead, the prospects for deep learning in personalized ECG diagnosis are undeniably promising. The future trajectory of this field is likely to be shaped by several key developments. Paramount among these is the enhancement of data collection efforts and the refinement of annotation practices. High-quality, diverse datasets are critical for training robust deep-learning models that can generalize well across different patient populations. Furthermore, the exploration and development of novel deep learning architectures, specifically tailored to capture the intricate and nuanced patterns within ECG signals, will be essential.

Improving model interpretability remains a vital objective to build trust and acceptance within clinical environments. Clinicians need to understand and trust the outputs of AI systems to integrate them effectively into their diagnostic workflows \cite{alahmadi2021explainable}. To this end, research into explainable AI (XAI) methods that elucidate the decision-making processes of deep learning models will be instrumental. Privacy-preserving machine learning techniques, such as federated learning and differential privacy, will play a crucial role in ensuring patient data is handled with the utmost care and confidentiality. These techniques will enable the development of robust AI models without compromising patient privacy, thus addressing one of the major ethical concerns in the deployment of AI in healthcare.

Collaboration with regulatory bodies will be essential to establish a comprehensive framework for the ethical and compliant deployment of AI technologies in healthcare settings \cite{ma2024evolution}. This collaboration will facilitate the development of guidelines and standards that ensure the safety, efficacy, and fairness of AI-based diagnostic tools. The design and development of user-centric systems that prioritize user experience and seamless clinical integration will be crucial for the widespread adoption of these technologies. Healthcare professionals must find these systems intuitive and beneficial to their workflows, which will encourage their acceptance and routine use.

Lastly, fostering cross-disciplinary collaboration will be key to creating solutions that are both technically advanced and clinically relevant \cite{topol2019high}. The integration of insights from cardiologists, data scientists, and engineers will be vital in developing AI systems that address real-world clinical needs and challenges \cite{dai2024deep, li2024integrated}.

In conclusion, while the journey toward integrating deep learning into personalized ECG diagnosis is replete with challenges, the collective efforts of the scientific, medical, and engineering communities have the potential to overcome these obstacles. The future of AI in cardiac care is poised to revolutionize diagnostic capabilities, enhance patient outcomes, and harmonize technology with the nuanced art of medicine, ultimately leading to a new era of precision cardiology \cite{khera2024ai}.
\section*{Acknowledgements}
This work was supported by the 

\section*{Competing interests}
The authors declare no competing interests.
 
\bibliography{reference}


\begin{thebibliography}{135}
\ifx \bisbn   \undefined \def \bisbn  #1{ISBN #1}\fi
\ifx \binits  \undefined \def \binits#1{#1}\fi
\ifx \bauthor  \undefined \def \bauthor#1{#1}\fi
\ifx \batitle  \undefined \def \batitle#1{#1}\fi
\ifx \bjtitle  \undefined \def \bjtitle#1{#1}\fi
\ifx \bvolume  \undefined \def \bvolume#1{\textbf{#1}}\fi
\ifx \byear  \undefined \def \byear#1{#1}\fi
\ifx \bissue  \undefined \def \bissue#1{#1}\fi
\ifx \bfpage  \undefined \def \bfpage#1{#1}\fi
\ifx \blpage  \undefined \def \blpage #1{#1}\fi
\ifx \burl  \undefined \def \burl#1{\textsf{#1}}\fi
\ifx \doiurl  \undefined \def \doiurl#1{\url{https://doi.org/#1}}\fi
\ifx \betal  \undefined \def \betal{\textit{et al.}}\fi
\ifx \binstitute  \undefined \def \binstitute#1{#1}\fi
\ifx \binstitutionaled  \undefined \def \binstitutionaled#1{#1}\fi
\ifx \bctitle  \undefined \def \bctitle#1{#1}\fi
\ifx \beditor  \undefined \def \beditor#1{#1}\fi
\ifx \bpublisher  \undefined \def \bpublisher#1{#1}\fi
\ifx \bbtitle  \undefined \def \bbtitle#1{#1}\fi
\ifx \bedition  \undefined \def \bedition#1{#1}\fi
\ifx \bseriesno  \undefined \def \bseriesno#1{#1}\fi
\ifx \blocation  \undefined \def \blocation#1{#1}\fi
\ifx \bsertitle  \undefined \def \bsertitle#1{#1}\fi
\ifx \bsnm \undefined \def \bsnm#1{#1}\fi
\ifx \bsuffix \undefined \def \bsuffix#1{#1}\fi
\ifx \bparticle \undefined \def \bparticle#1{#1}\fi
\ifx \barticle \undefined \def \barticle#1{#1}\fi
\bibcommenthead
\ifx \bconfdate \undefined \def \bconfdate #1{#1}\fi
\ifx \botherref \undefined \def \botherref #1{#1}\fi
\ifx \url \undefined \def \url#1{\textsf{#1}}\fi
\ifx \bchapter \undefined \def \bchapter#1{#1}\fi
\ifx \bbook \undefined \def \bbook#1{#1}\fi
\ifx \bcomment \undefined \def \bcomment#1{#1}\fi
\ifx \oauthor \undefined \def \oauthor#1{#1}\fi
\ifx \citeauthoryear \undefined \def \citeauthoryear#1{#1}\fi
\ifx \endbibitem  \undefined \def \endbibitem {}\fi
\ifx \bconflocation  \undefined \def \bconflocation#1{#1}\fi
\ifx \arxivurl  \undefined \def \arxivurl#1{\textsf{#1}}\fi
\csname PreBibitemsHook\endcsname

\bibitem[\protect\citeauthoryear{Bhatia and Dorian}{2018}]{bhatia2018screening}
\begin{barticle}
\bauthor{\bsnm{Bhatia}, \binits{R.S.}},
\bauthor{\bsnm{Dorian}, \binits{P.}}:
\batitle{Screening for cardiovascular disease risk with electrocardiography}.
\bjtitle{JAMA Internal Medicine}
\bvolume{178}(\bissue{9}),
\bfpage{1163}--\blpage{1164}
(\byear{2018})
\end{barticle}
\endbibitem

\bibitem[\protect\citeauthoryear{Martis et~al.}{2014}]{martis2014current}
\begin{barticle}
\bauthor{\bsnm{Martis}, \binits{R.J.}},
\bauthor{\bsnm{Acharya}, \binits{U.R.}},
\bauthor{\bsnm{Adeli}, \binits{H.}}:
\batitle{Current methods in electrocardiogram characterization}.
\bjtitle{Computers in biology and medicine}
\bvolume{48},
\bfpage{133}--\blpage{149}
(\byear{2014})
\end{barticle}
\endbibitem

\bibitem[\protect\citeauthoryear{Liu et~al.}{2021}]{liu2021deep}
\begin{barticle}
\bauthor{\bsnm{Liu}, \binits{X.}},
\bauthor{\bsnm{Wang}, \binits{H.}},
\bauthor{\bsnm{Li}, \binits{Z.}},
\bauthor{\bsnm{Qin}, \binits{L.}}:
\batitle{Deep learning in ecg diagnosis: A review}.
\bjtitle{Knowledge-Based Systems}
\bvolume{227},
\bfpage{107187}
(\byear{2021})
\end{barticle}
\endbibitem

\bibitem[\protect\citeauthoryear{Rafie et~al.}{2021}]{rafie2021ecg}
\begin{barticle}
\bauthor{\bsnm{Rafie}, \binits{N.}},
\bauthor{\bsnm{Kashou}, \binits{A.H.}},
\bauthor{\bsnm{Noseworthy}, \binits{P.A.}}:
\batitle{Ecg interpretation: clinical relevance, challenges, and advances}.
\bjtitle{Hearts}
\bvolume{2}(\bissue{4}),
\bfpage{505}--\blpage{513}
(\byear{2021})
\end{barticle}
\endbibitem

\bibitem[\protect\citeauthoryear{Khera et~al.}{2024}]{khera2024transforming}
\begin{barticle}
\bauthor{\bsnm{Khera}, \binits{R.}},
\bauthor{\bsnm{Oikonomou}, \binits{E.K.}},
\bauthor{\bsnm{Nadkarni}, \binits{G.N.}},
\bauthor{\bsnm{Morley}, \binits{J.R.}},
\bauthor{\bsnm{Wiens}, \binits{J.}},
\bauthor{\bsnm{Butte}, \binits{A.J.}},
\bauthor{\bsnm{Topol}, \binits{E.J.}}:
\batitle{Transforming cardiovascular care with artificial intelligence: From discovery to practice: Jacc state-of-the-art review}.
\bjtitle{Journal of the American College of Cardiology}
\bvolume{84}(\bissue{1}),
\bfpage{97}--\blpage{114}
(\byear{2024})
\end{barticle}
\endbibitem

\bibitem[\protect\citeauthoryear{Ni et~al.}{2024}]{ni2024survey}
\begin{botherref}
\oauthor{\bsnm{Ni}, \binits{J.}},
\oauthor{\bsnm{Tang}, \binits{H.}},
\oauthor{\bsnm{Haque}, \binits{S.T.}},
\oauthor{\bsnm{Yan}, \binits{Y.}},
\oauthor{\bsnm{Ngu}, \binits{A.H.}}:
A survey on multimodal wearable sensor-based human action recognition.
arXiv preprint arXiv:2404.15349
(2024)
\end{botherref}
\endbibitem

\bibitem[\protect\citeauthoryear{Hammad et~al.}{2018}]{hammad2018detection}
\begin{barticle}
\bauthor{\bsnm{Hammad}, \binits{M.}},
\bauthor{\bsnm{Maher}, \binits{A.}},
\bauthor{\bsnm{Wang}, \binits{K.}},
\bauthor{\bsnm{Jiang}, \binits{F.}},
\bauthor{\bsnm{Amrani}, \binits{M.}}:
\batitle{Detection of abnormal heart conditions based on characteristics of ecg signals}.
\bjtitle{Measurement}
\bvolume{125},
\bfpage{634}--\blpage{644}
(\byear{2018})
\end{barticle}
\endbibitem

\bibitem[\protect\citeauthoryear{Berkaya et~al.}{2018}]{berkaya2018survey}
\begin{barticle}
\bauthor{\bsnm{Berkaya}, \binits{S.K.}},
\bauthor{\bsnm{Uysal}, \binits{A.K.}},
\bauthor{\bsnm{Gunal}, \binits{E.S.}},
\bauthor{\bsnm{Ergin}, \binits{S.}},
\bauthor{\bsnm{Gunal}, \binits{S.}},
\bauthor{\bsnm{Gulmezoglu}, \binits{M.B.}}:
\batitle{A survey on ecg analysis}.
\bjtitle{Biomedical Signal Processing and Control}
\bvolume{43},
\bfpage{216}--\blpage{235}
(\byear{2018})
\end{barticle}
\endbibitem

\bibitem[\protect\citeauthoryear{Ribeiro et~al.}{2020}]{ribeiro2020automatic}
\begin{barticle}
\bauthor{\bsnm{Ribeiro}, \binits{A.H.}},
\bauthor{\bsnm{Ribeiro}, \binits{M.H.}},
\bauthor{\bsnm{Paix{\~a}o}, \binits{G.M.}},
\bauthor{\bsnm{Oliveira}, \binits{D.M.}},
\bauthor{\bsnm{Gomes}, \binits{P.R.}},
\bauthor{\bsnm{Canazart}, \binits{J.A.}},
\bauthor{\bsnm{Ferreira}, \binits{M.P.}},
\bauthor{\bsnm{Andersson}, \binits{C.R.}},
\bauthor{\bsnm{Macfarlane}, \binits{P.W.}},
\bauthor{\bsnm{Meira~Jr}, \binits{W.}}, \betal:
\batitle{Automatic diagnosis of the 12-lead ecg using a deep neural network}.
\bjtitle{Nature communications}
\bvolume{11}(\bissue{1}),
\bfpage{1760}
(\byear{2020})
\end{barticle}
\endbibitem

\bibitem[\protect\citeauthoryear{Fratini et~al.}{2015}]{fratini2015individual}
\begin{barticle}
\bauthor{\bsnm{Fratini}, \binits{A.}},
\bauthor{\bsnm{Sansone}, \binits{M.}},
\bauthor{\bsnm{Bifulco}, \binits{P.}},
\bauthor{\bsnm{Cesarelli}, \binits{M.}}:
\batitle{Individual identification via electrocardiogram analysis}.
\bjtitle{Biomedical engineering online}
\bvolume{14},
\bfpage{1}--\blpage{23}
(\byear{2015})
\end{barticle}
\endbibitem

\bibitem[\protect\citeauthoryear{Anbalagan et~al.}{2023}]{anbalagan2023analysis}
\begin{botherref}
\oauthor{\bsnm{Anbalagan}, \binits{T.}},
\oauthor{\bsnm{Nath}, \binits{M.K.}},
\oauthor{\bsnm{Vijayalakshmi}, \binits{D.}},
\oauthor{\bsnm{Anbalagan}, \binits{A.}}:
Analysis of various techniques for ecg signal in healthcare, past, present, and future.
Biomedical Engineering Advances,
100089
(2023)
\end{botherref}
\endbibitem

\bibitem[\protect\citeauthoryear{Li et~al.}{2018}]{li2018patient}
\begin{barticle}
\bauthor{\bsnm{Li}, \binits{Y.}},
\bauthor{\bsnm{Pang}, \binits{Y.}},
\bauthor{\bsnm{Wang}, \binits{J.}},
\bauthor{\bsnm{Li}, \binits{X.}}:
\batitle{Patient-specific ecg classification by deeper cnn from generic to dedicated}.
\bjtitle{Neurocomputing}
\bvolume{314},
\bfpage{336}--\blpage{346}
(\byear{2018})
\end{barticle}
\endbibitem

\bibitem[\protect\citeauthoryear{Ebrahimi et~al.}{2020}]{ebrahimi2020review}
\begin{barticle}
\bauthor{\bsnm{Ebrahimi}, \binits{Z.}},
\bauthor{\bsnm{Loni}, \binits{M.}},
\bauthor{\bsnm{Daneshtalab}, \binits{M.}},
\bauthor{\bsnm{Gharehbaghi}, \binits{A.}}:
\batitle{A review on deep learning methods for ecg arrhythmia classification}.
\bjtitle{Expert Systems with Applications: X}
\bvolume{7},
\bfpage{100033}
(\byear{2020})
\end{barticle}
\endbibitem

\bibitem[\protect\citeauthoryear{Siontis et~al.}{2021}]{siontis2021artificial}
\begin{barticle}
\bauthor{\bsnm{Siontis}, \binits{K.C.}},
\bauthor{\bsnm{Noseworthy}, \binits{P.A.}},
\bauthor{\bsnm{Attia}, \binits{Z.I.}},
\bauthor{\bsnm{Friedman}, \binits{P.A.}}:
\batitle{Artificial intelligence-enhanced electrocardiography in cardiovascular disease management}.
\bjtitle{Nature Reviews Cardiology}
\bvolume{18}(\bissue{7}),
\bfpage{465}--\blpage{478}
(\byear{2021})
\end{barticle}
\endbibitem

\bibitem[\protect\citeauthoryear{Weimann and Conrad}{2021}]{weimann2021transfer}
\begin{barticle}
\bauthor{\bsnm{Weimann}, \binits{K.}},
\bauthor{\bsnm{Conrad}, \binits{T.O.}}:
\batitle{Transfer learning for ecg classification}.
\bjtitle{Scientific reports}
\bvolume{11}(\bissue{1}),
\bfpage{5251}
(\byear{2021})
\end{barticle}
\endbibitem

\bibitem[\protect\citeauthoryear{Pal et~al.}{2021}]{pal2021cardionet}
\begin{barticle}
\bauthor{\bsnm{Pal}, \binits{A.}},
\bauthor{\bsnm{Srivastva}, \binits{R.}},
\bauthor{\bsnm{Singh}, \binits{Y.N.}}:
\batitle{Cardionet: An efficient ecg arrhythmia classification system using transfer learning}.
\bjtitle{Big Data Research}
\bvolume{26},
\bfpage{100271}
(\byear{2021})
\end{barticle}
\endbibitem

\bibitem[\protect\citeauthoryear{Maray et~al.}{2023}]{maray2023transfer}
\begin{barticle}
\bauthor{\bsnm{Maray}, \binits{N.}},
\bauthor{\bsnm{Ngu}, \binits{A.H.}},
\bauthor{\bsnm{Ni}, \binits{J.}},
\bauthor{\bsnm{Debnath}, \binits{M.}},
\bauthor{\bsnm{Wang}, \binits{L.}}:
\batitle{Transfer learning on small datasets for improved fall detection}.
\bjtitle{Sensors}
\bvolume{23}(\bissue{3}),
\bfpage{1105}
(\byear{2023})
\end{barticle}
\endbibitem

\bibitem[\protect\citeauthoryear{Mohebbanaaz et~al.}{2022}]{mohebbanaaz2022new}
\begin{barticle}
\bauthor{\bsnm{Mohebbanaaz}},
\bauthor{\bsnm{Kumar}, \binits{L.R.}},
\bauthor{\bsnm{Sai}, \binits{Y.P.}}:
\batitle{A new transfer learning approach to detect cardiac arrhythmia from ecg signals}.
\bjtitle{Signal, Image and Video Processing}
\bvolume{16}(\bissue{7}),
\bfpage{1945}--\blpage{1953}
(\byear{2022})
\end{barticle}
\endbibitem

\bibitem[\protect\citeauthoryear{Ng et~al.}{2023}]{ng2023few}
\begin{barticle}
\bauthor{\bsnm{Ng}, \binits{Y.}},
\bauthor{\bsnm{Liao}, \binits{M.-T.}},
\bauthor{\bsnm{Chen}, \binits{T.-L.}},
\bauthor{\bsnm{Lee}, \binits{C.-K.}},
\bauthor{\bsnm{Chou}, \binits{C.-Y.}},
\bauthor{\bsnm{Wang}, \binits{W.}}:
\batitle{Few-shot transfer learning for personalized atrial fibrillation detection using patient-based siamese network with single-lead ecg records}.
\bjtitle{Artificial Intelligence in Medicine}
\bvolume{144},
\bfpage{102644}
(\byear{2023})
\end{barticle}
\endbibitem

\bibitem[\protect\citeauthoryear{Hu et~al.}{2023a}]{hu2023personalized}
\begin{botherref}
\oauthor{\bsnm{Hu}, \binits{S.}},
\oauthor{\bsnm{Wang}, \binits{Y.}},
\oauthor{\bsnm{Liu}, \binits{J.}},
\oauthor{\bsnm{Yang}, \binits{C.}}:
Personalized transfer learning for single-lead ecg-based sleep apnea detection: Exploring the label mapping length and transfer strategy using hybrid transformer model.
IEEE Transactions on Instrumentation and Measurement
(2023)
\end{botherref}
\endbibitem

\bibitem[\protect\citeauthoryear{Hu et~al.}{2023b}]{hu2023semi}
\begin{botherref}
\oauthor{\bsnm{Hu}, \binits{S.}},
\oauthor{\bsnm{Liu}, \binits{J.}},
\oauthor{\bsnm{Yang}, \binits{C.}},
\oauthor{\bsnm{Wang}, \binits{A.}},
\oauthor{\bsnm{Li}, \binits{K.}},
\oauthor{\bsnm{Liu}, \binits{W.}}, et al.:
Semi-supervised learning for low-cost personalized obstructive sleep apnea detection using unsupervised deep learning and single-lead electrocardiogram.
IEEE Journal of Biomedical and Health Informatics
(2023)
\end{botherref}
\endbibitem

\bibitem[\protect\citeauthoryear{Hong et~al.}{2021}]{hong2021deep}
\begin{bchapter}
\bauthor{\bsnm{Hong}, \binits{J.}},
\bauthor{\bsnm{Gao}, \binits{J.}},
\bauthor{\bsnm{Liu}, \binits{Q.}},
\bauthor{\bsnm{Zhang}, \binits{Y.}},
\bauthor{\bsnm{Zheng}, \binits{Y.}}:
\bctitle{Deep learning model with individualized fine-tuning for dynamic and beat-to-beat blood pressure estimation}.
In: \bbtitle{2021 IEEE 17th International Conference on Wearable and Implantable Body Sensor Networks (BSN)},
pp. \bfpage{1}--\blpage{4}
(\byear{2021}).
\bcomment{IEEE}
\end{bchapter}
\endbibitem

\bibitem[\protect\citeauthoryear{Suhas et~al.}{2024}]{suhas2024end}
\begin{bchapter}
\bauthor{\bsnm{Suhas}, \binits{B.}},
\bauthor{\bsnm{Srinivasa}, \binits{R.S.}},
\bauthor{\bsnm{Saidutta}, \binits{Y.M.}},
\bauthor{\bsnm{Cho}, \binits{J.}},
\bauthor{\bsnm{Lee}, \binits{C.-H.}},
\bauthor{\bsnm{Yang}, \binits{C.}},
\bauthor{\bsnm{Shen}, \binits{Y.}},
\bauthor{\bsnm{Jin}, \binits{H.}}:
\bctitle{End-to-end personalized cuff-less blood pressure monitoring using ecg and ppg signals}.
In: \bbtitle{ICASSP 2024-2024 IEEE International Conference on Acoustics, Speech and Signal Processing (ICASSP)},
pp. \bfpage{2101}--\blpage{2105}
(\byear{2024}).
\bcomment{IEEE}
\end{bchapter}
\endbibitem

\bibitem[\protect\citeauthoryear{Liu et~al.}{2024}]{liu2024etp}
\begin{bchapter}
\bauthor{\bsnm{Liu}, \binits{C.}},
\bauthor{\bsnm{Wan}, \binits{Z.}},
\bauthor{\bsnm{Cheng}, \binits{S.}},
\bauthor{\bsnm{Zhang}, \binits{M.}},
\bauthor{\bsnm{Arcucci}, \binits{R.}}:
\bctitle{Etp: Learning transferable ecg representations via ecg-text pre-training}.
In: \bbtitle{ICASSP 2024-2024 IEEE International Conference on Acoustics, Speech and Signal Processing (ICASSP)},
pp. \bfpage{8230}--\blpage{8234}
(\byear{2024}).
\bcomment{IEEE}
\end{bchapter}
\endbibitem

\bibitem[\protect\citeauthoryear{Jia et~al.}{2021a}]{jia2021learning}
\begin{bchapter}
\bauthor{\bsnm{Jia}, \binits{Z.}},
\bauthor{\bsnm{Wang}, \binits{Z.}},
\bauthor{\bsnm{Hong}, \binits{F.}},
\bauthor{\bsnm{Ping}, \binits{L.}},
\bauthor{\bsnm{Shi}, \binits{Y.}},
\bauthor{\bsnm{Hu}, \binits{J.}}:
\bctitle{Learning to learn personalized neural network for ventricular arrhythmias detection on intracardiac egms.}
In: \bbtitle{IJCAI},
pp. \bfpage{2606}--\blpage{2613}
(\byear{2021})
\end{bchapter}
\endbibitem

\bibitem[\protect\citeauthoryear{Jia et~al.}{2021b}]{jia2021device}
\begin{barticle}
\bauthor{\bsnm{Jia}, \binits{Z.}},
\bauthor{\bsnm{Shi}, \binits{Y.}},
\bauthor{\bsnm{Saba}, \binits{S.}},
\bauthor{\bsnm{Hu}, \binits{J.}}:
\batitle{On-device prior knowledge incorporated learning for personalized atrial fibrillation detection}.
\bjtitle{ACM Transactions on Embedded Computing Systems (TECS)}
\bvolume{20}(\bissue{5s}),
\bfpage{1}--\blpage{25}
(\byear{2021})
\end{barticle}
\endbibitem

\bibitem[\protect\citeauthoryear{Ullah et~al.}{2022}]{ullah2022automatic}
\begin{barticle}
\bauthor{\bsnm{Ullah}, \binits{H.}},
\bauthor{\bsnm{Heyat}, \binits{M.B.B.}},
\bauthor{\bsnm{Akhtar}, \binits{F.}},
\bauthor{\bsnm{Muaad}, \binits{A.Y.}},
\bauthor{\bsnm{Ukwuoma}, \binits{C.C.}},
\bauthor{\bsnm{Bilal}, \binits{M.}},
\bauthor{\bsnm{Miraz}, \binits{M.H.}},
\bauthor{\bsnm{Bhuiyan}, \binits{M.A.S.}},
\bauthor{\bsnm{Wu}, \binits{K.}},
\bauthor{\bsnm{Dama{\v{s}}evi{\v{c}}ius}, \binits{R.}}, \betal:
\batitle{An automatic premature ventricular contraction recognition system based on imbalanced dataset and pre-trained residual network using transfer learning on ecg signal}.
\bjtitle{Diagnostics}
\bvolume{13}(\bissue{1}),
\bfpage{87}
(\byear{2022})
\end{barticle}
\endbibitem

\bibitem[\protect\citeauthoryear{Ma et~al.}{2023}]{ma2023atrial}
\begin{botherref}
\oauthor{\bsnm{Ma}, \binits{C.}},
\oauthor{\bsnm{Long}, \binits{X.}},
\oauthor{\bsnm{Sheng}, \binits{W.}},
\oauthor{\bsnm{Vullings}, \binits{R.}},
\oauthor{\bsnm{Yang}, \binits{M.}},
\oauthor{\bsnm{Zhao}, \binits{L.}},
\oauthor{\bsnm{Aarts}, \binits{R.M.}},
\oauthor{\bsnm{Li}, \binits{J.}},
\oauthor{\bsnm{Liu}, \binits{C.}}:
An atrial fibrillation detection strategy in dynamic ecgs with significant individual differences.
IEEE Transactions on Instrumentation and Measurement
(2023)
\end{botherref}
\endbibitem

\bibitem[\protect\citeauthoryear{He et~al.}{2021}]{he2021online}
\begin{bchapter}
\bauthor{\bsnm{He}, \binits{W.}},
\bauthor{\bsnm{Ye}, \binits{Y.}},
\bauthor{\bsnm{Li}, \binits{Y.}},
\bauthor{\bsnm{Pan}, \binits{T.}},
\bauthor{\bsnm{Lu}, \binits{L.}}:
\bctitle{Online cross-subject emotion recognition from ecg via unsupervised domain adaptation}.
In: \bbtitle{2021 43rd Annual International Conference of the IEEE Engineering in Medicine \& Biology Society (EMBC)},
pp. \bfpage{1001}--\blpage{1005}
(\byear{2021}).
\bcomment{IEEE}
\end{bchapter}
\endbibitem

\bibitem[\protect\citeauthoryear{Wang et~al.}{2021}]{wang2021inter}
\begin{barticle}
\bauthor{\bsnm{Wang}, \binits{G.}},
\bauthor{\bsnm{Chen}, \binits{M.}},
\bauthor{\bsnm{Ding}, \binits{Z.}},
\bauthor{\bsnm{Li}, \binits{J.}},
\bauthor{\bsnm{Yang}, \binits{H.}},
\bauthor{\bsnm{Zhang}, \binits{P.}}:
\batitle{Inter-patient ecg arrhythmia heartbeat classification based on unsupervised domain adaptation}.
\bjtitle{Neurocomputing}
\bvolume{454},
\bfpage{339}--\blpage{349}
(\byear{2021})
\end{barticle}
\endbibitem

\bibitem[\protect\citeauthoryear{Chen et~al.}{2020}]{chen2020unsupervised}
\begin{bchapter}
\bauthor{\bsnm{Chen}, \binits{M.}},
\bauthor{\bsnm{Wang}, \binits{G.}},
\bauthor{\bsnm{Ding}, \binits{Z.}},
\bauthor{\bsnm{Li}, \binits{J.}},
\bauthor{\bsnm{Yang}, \binits{H.}}:
\bctitle{Unsupervised domain adaptation for ecg arrhythmia classification}.
In: \bbtitle{2020 42nd Annual International Conference of the IEEE Engineering in Medicine \& Biology Society (EMBC)},
pp. \bfpage{304}--\blpage{307}
(\byear{2020}).
\bcomment{IEEE}
\end{bchapter}
\endbibitem

\bibitem[\protect\citeauthoryear{Yuan and Siyal}{2023}]{yuan2023target}
\begin{barticle}
\bauthor{\bsnm{Yuan}, \binits{L.}},
\bauthor{\bsnm{Siyal}, \binits{M.Y.}}:
\batitle{Target-oriented augmentation privacy-protection domain adaptation for imbalanced ecg beat classification}.
\bjtitle{Biomedical Signal Processing and Control}
\bvolume{86},
\bfpage{105308}
(\byear{2023})
\end{barticle}
\endbibitem

\bibitem[\protect\citeauthoryear{Du et~al.}{2023}]{du2023diagnosis}
\begin{barticle}
\bauthor{\bsnm{Du}, \binits{M.}},
\bauthor{\bsnm{Yang}, \binits{Y.}},
\bauthor{\bsnm{Zhang}, \binits{L.}}:
\batitle{Diagnosis of atrial fibrillation based on unsupervised domain adaptation}.
\bjtitle{Computers in Biology and Medicine}
\bvolume{164},
\bfpage{107275}
(\byear{2023})
\end{barticle}
\endbibitem

\bibitem[\protect\citeauthoryear{Feng et~al.}{2022}]{feng2022unsupervised}
\begin{barticle}
\bauthor{\bsnm{Feng}, \binits{P.}},
\bauthor{\bsnm{Fu}, \binits{J.}},
\bauthor{\bsnm{Ge}, \binits{Z.}},
\bauthor{\bsnm{Wang}, \binits{H.}},
\bauthor{\bsnm{Zhou}, \binits{Y.}},
\bauthor{\bsnm{Zhou}, \binits{B.}},
\bauthor{\bsnm{Wang}, \binits{Z.}}:
\batitle{Unsupervised semantic-aware adaptive feature fusion network for arrhythmia detection}.
\bjtitle{Information Sciences}
\bvolume{582},
\bfpage{509}--\blpage{528}
(\byear{2022})
\end{barticle}
\endbibitem

\bibitem[\protect\citeauthoryear{Golany and Radinsky}{2019}]{golany2019pgans}
\begin{bchapter}
\bauthor{\bsnm{Golany}, \binits{T.}},
\bauthor{\bsnm{Radinsky}, \binits{K.}}:
\bctitle{Pgans: Personalized generative adversarial networks for ecg synthesis to improve patient-specific deep ecg classification}.
In: \bbtitle{Proceedings of the AAAI Conference on Artificial Intelligence},
vol. \bseriesno{33},
pp. \bfpage{557}--\blpage{564}
(\byear{2019})
\end{bchapter}
\endbibitem

\bibitem[\protect\citeauthoryear{Delaney et~al.}{2019}]{delaney2019synthesis}
\begin{botherref}
\oauthor{\bsnm{Delaney}, \binits{A.M.}},
\oauthor{\bsnm{Brophy}, \binits{E.}},
\oauthor{\bsnm{Ward}, \binits{T.E.}}:
Synthesis of Realistic ECG using Generative Adversarial Networks
(2019)
\end{botherref}
\endbibitem

\bibitem[\protect\citeauthoryear{Shaker et~al.}{2020}]{shaker2020generalization}
\begin{barticle}
\bauthor{\bsnm{Shaker}, \binits{A.M.}},
\bauthor{\bsnm{Tantawi}, \binits{M.}},
\bauthor{\bsnm{Shedeed}, \binits{H.A.}},
\bauthor{\bsnm{Tolba}, \binits{M.F.}}:
\batitle{Generalization of convolutional neural networks for ecg classification using generative adversarial networks}.
\bjtitle{IEEE Access}
\bvolume{8},
\bfpage{35592}--\blpage{35605}
(\byear{2020})
\end{barticle}
\endbibitem

\bibitem[\protect\citeauthoryear{Chen et~al.}{2022}]{chen2022me}
\begin{bchapter}
\bauthor{\bsnm{Chen}, \binits{J.}},
\bauthor{\bsnm{Liao}, \binits{K.}},
\bauthor{\bsnm{Wei}, \binits{K.}},
\bauthor{\bsnm{Ying}, \binits{H.}},
\bauthor{\bsnm{Chen}, \binits{D.Z.}},
\bauthor{\bsnm{Wu}, \binits{J.}}:
\bctitle{Me-gan: Learning panoptic electrocardio representations for multi-view ecg synthesis conditioned on heart diseases}.
In: \bbtitle{International Conference on Machine Learning},
pp. \bfpage{3360}--\blpage{3370}
(\byear{2022}).
\bcomment{PMLR}
\end{bchapter}
\endbibitem

\bibitem[\protect\citeauthoryear{Golany et~al.}{2021}]{golany2021ecg}
\begin{botherref}
\oauthor{\bsnm{Golany}, \binits{T.}},
\oauthor{\bsnm{Freedman}, \binits{D.}},
\oauthor{\bsnm{Radinsky}, \binits{K.}}:
Ecg ode-gan: Learning ordinary differential equations of ecg dynamics via generative adversarial learning
\textbf{35}(1),
134--141
(2021)
\end{botherref}
\endbibitem

\bibitem[\protect\citeauthoryear{Neifar et~al.}{2024}]{neifar2024leveraging}
\begin{botherref}
\oauthor{\bsnm{Neifar}, \binits{N.}},
\oauthor{\bsnm{Ben-Hamadou}, \binits{A.}},
\oauthor{\bsnm{Mdhaffar}, \binits{A.}},
\oauthor{\bsnm{Jmaiel}, \binits{M.}},
\oauthor{\bsnm{Freisleben}, \binits{B.}}:
Leveraging statistical shape priors in gan-based ecg synthesis.
IEEE Access
(2024)
\end{botherref}
\endbibitem

\bibitem[\protect\citeauthoryear{Sarkar and Etemad}{2021}]{sarkar2021cardiogan}
\begin{bchapter}
\bauthor{\bsnm{Sarkar}, \binits{P.}},
\bauthor{\bsnm{Etemad}, \binits{A.}}:
\bctitle{Cardiogan: Attentive generative adversarial network with dual discriminators for synthesis of ecg from ppg}.
In: \bbtitle{Proceedings of the AAAI Conference on Artificial Intelligence},
vol. \bseriesno{35},
pp. \bfpage{488}--\blpage{496}
(\byear{2021})
\end{bchapter}
\endbibitem

\bibitem[\protect\citeauthoryear{Kim and Pan}{2021}]{kim2021study}
\begin{barticle}
\bauthor{\bsnm{Kim}, \binits{M.-G.}},
\bauthor{\bsnm{Pan}, \binits{S.B.}}:
\batitle{A study on user recognition using the generated synthetic electrocardiogram signal}.
\bjtitle{Sensors}
\bvolume{21}(\bissue{5}),
\bfpage{1887}
(\byear{2021})
\end{barticle}
\endbibitem

\bibitem[\protect\citeauthoryear{Golany et~al.}{2020}]{golany2020simgans}
\begin{bchapter}
\bauthor{\bsnm{Golany}, \binits{T.}},
\bauthor{\bsnm{Radinsky}, \binits{K.}},
\bauthor{\bsnm{Freedman}, \binits{D.}}:
\bctitle{Simgans: Simulator-based generative adversarial networks for ecg synthesis to improve deep ecg classification}.
In: \bbtitle{International Conference on Machine Learning},
pp. \bfpage{3597}--\blpage{3606}
(\byear{2020}).
\bcomment{PMLR}
\end{bchapter}
\endbibitem

\bibitem[\protect\citeauthoryear{Kang et~al.}{2023}]{kang2023gan}
\begin{barticle}
\bauthor{\bsnm{Kang}, \binits{Y.}},
\bauthor{\bsnm{Yang}, \binits{G.}},
\bauthor{\bsnm{Eom}, \binits{H.}},
\bauthor{\bsnm{Han}, \binits{S.}},
\bauthor{\bsnm{Baek}, \binits{S.}},
\bauthor{\bsnm{Noh}, \binits{S.}},
\bauthor{\bsnm{Shin}, \binits{Y.}},
\bauthor{\bsnm{Park}, \binits{C.}}:
\batitle{Gan-based patient information hiding for an ecg authentication system}.
\bjtitle{Biomedical Engineering Letters}
\bvolume{13}(\bissue{2}),
\bfpage{197}--\blpage{207}
(\byear{2023})
\end{barticle}
\endbibitem

\bibitem[\protect\citeauthoryear{Wang et~al.}{2023}]{wang2023hierarchical}
\begin{barticle}
\bauthor{\bsnm{Wang}, \binits{Z.}},
\bauthor{\bsnm{Stavrakis}, \binits{S.}},
\bauthor{\bsnm{Yao}, \binits{B.}}:
\batitle{Hierarchical deep learning with generative adversarial network for automatic cardiac diagnosis from ecg signals}.
\bjtitle{Computers in Biology and Medicine}
\bvolume{155},
\bfpage{106641}
(\byear{2023})
\end{barticle}
\endbibitem

\bibitem[\protect\citeauthoryear{Byeon and Kwak}{2023}]{byeon2023semi}
\begin{barticle}
\bauthor{\bsnm{Byeon}, \binits{Y.-H.}},
\bauthor{\bsnm{Kwak}, \binits{K.-C.}}:
\batitle{Semi-supervised domain adaptation for individual identification from electrocardiogram signals}.
\bjtitle{Applied Sciences}
\bvolume{13}(\bissue{24}),
\bfpage{13259}
(\byear{2023})
\end{barticle}
\endbibitem

\bibitem[\protect\citeauthoryear{Rafi and Ko}{2023}]{rafi2023sf}
\begin{barticle}
\bauthor{\bsnm{Rafi}, \binits{T.H.}},
\bauthor{\bsnm{Ko}, \binits{Y.-W.}}:
\batitle{Sf-ecg: Source-free intersubject domain adaptation for electrocardiography-based arrhythmia classification}.
\bjtitle{Applied Sciences}
\bvolume{13}(\bissue{14}),
\bfpage{8551}
(\byear{2023})
\end{barticle}
\endbibitem

\bibitem[\protect\citeauthoryear{Vo et~al.}{2021}]{vo2021p2e}
\begin{bchapter}
\bauthor{\bsnm{Vo}, \binits{K.}},
\bauthor{\bsnm{Naeini}, \binits{E.K.}},
\bauthor{\bsnm{Naderi}, \binits{A.}},
\bauthor{\bsnm{Jilani}, \binits{D.}},
\bauthor{\bsnm{Rahmani}, \binits{A.M.}},
\bauthor{\bsnm{Dutt}, \binits{N.}},
\bauthor{\bsnm{Cao}, \binits{H.}}:
\bctitle{P2e-wgan: Ecg waveform synthesis from ppg with conditional wasserstein generative adversarial networks}.
In: \bbtitle{Proceedings of the 36th Annual ACM Symposium on Applied Computing},
pp. \bfpage{1030}--\blpage{1036}
(\byear{2021})
\end{bchapter}
\endbibitem

\bibitem[\protect\citeauthoryear{Joo et~al.}{2023}]{joo2023twelve}
\begin{bchapter}
\bauthor{\bsnm{Joo}, \binits{J.}},
\bauthor{\bsnm{Joo}, \binits{G.}},
\bauthor{\bsnm{Kim}, \binits{Y.}},
\bauthor{\bsnm{Jin}, \binits{M.-N.}},
\bauthor{\bsnm{Park}, \binits{J.}},
\bauthor{\bsnm{Im}, \binits{H.}}:
\bctitle{Twelve-lead ecg reconstruction from single-lead signals using generative adversarial networks}.
In: \bbtitle{International Conference on Medical Image Computing and Computer-Assisted Intervention},
pp. \bfpage{184}--\blpage{194}
(\byear{2023}).
\bcomment{Springer}
\end{bchapter}
\endbibitem

\bibitem[\protect\citeauthoryear{Kuo et~al.}{2022}]{kuo2022towards}
\begin{barticle}
\bauthor{\bsnm{Kuo}, \binits{C.-E.}},
\bauthor{\bsnm{Lu}, \binits{T.-H.}},
\bauthor{\bsnm{Chen}, \binits{G.-T.}},
\bauthor{\bsnm{Liao}, \binits{P.-Y.}}:
\batitle{Towards precision sleep medicine: Self-attention gan as an innovative data augmentation technique for developing personalized automatic sleep scoring classification}.
\bjtitle{Computers in Biology and Medicine}
\bvolume{148},
\bfpage{105828}
(\byear{2022})
\end{barticle}
\endbibitem

\bibitem[\protect\citeauthoryear{Ho et~al.}{2020}]{ho2020denoising}
\begin{barticle}
\bauthor{\bsnm{Ho}, \binits{J.}},
\bauthor{\bsnm{Jain}, \binits{A.}},
\bauthor{\bsnm{Abbeel}, \binits{P.}}:
\batitle{Denoising diffusion probabilistic models}.
\bjtitle{Advances in neural information processing systems}
\bvolume{33},
\bfpage{6840}--\blpage{6851}
(\byear{2020})
\end{barticle}
\endbibitem

\bibitem[\protect\citeauthoryear{Ouyang et~al.}{2024}]{ouyang2024transfer}
\begin{botherref}
\oauthor{\bsnm{Ouyang}, \binits{Y.}},
\oauthor{\bsnm{Xie}, \binits{L.}},
\oauthor{\bsnm{Zha}, \binits{H.}},
\oauthor{\bsnm{Cheng}, \binits{G.}}:
Transfer learning for diffusion models.
arXiv preprint arXiv:2405.16876
(2024)
\end{botherref}
\endbibitem

\bibitem[\protect\citeauthoryear{Debnath et~al.}{2024}]{debnath2024impact}
\begin{bchapter}
\bauthor{\bsnm{Debnath}, \binits{M.}},
\bauthor{\bsnm{Kabir}, \binits{M.S.}},
\bauthor{\bsnm{Ni}, \binits{J.}},
\bauthor{\bsnm{Ngu}, \binits{A.H.H.}}:
\bctitle{The impact of synthetic data on fall detection application}.
In: \bbtitle{International Conference on Artificial Intelligence in Medicine},
pp. \bfpage{204}--\blpage{209}
(\byear{2024}).
\bcomment{Springer}
\end{bchapter}
\endbibitem

\bibitem[\protect\citeauthoryear{Adib et~al.}{2023}]{adib2023synthetic}
\begin{botherref}
\oauthor{\bsnm{Adib}, \binits{E.}},
\oauthor{\bsnm{Fernandez}, \binits{A.S.}},
\oauthor{\bsnm{Afghah}, \binits{F.}},
\oauthor{\bsnm{Prevost}, \binits{J.J.}}:
Synthetic ecg signal generation using probabilistic diffusion models.
IEEE Access
(2023)
\end{botherref}
\endbibitem

\bibitem[\protect\citeauthoryear{Alcaraz and Strodthoff}{2023}]{alcaraz2023diffusion}
\begin{barticle}
\bauthor{\bsnm{Alcaraz}, \binits{J.M.L.}},
\bauthor{\bsnm{Strodthoff}, \binits{N.}}:
\batitle{Diffusion-based conditional ecg generation with structured state space models}.
\bjtitle{Computers in Biology and Medicine}
\bvolume{163},
\bfpage{107115}
(\byear{2023})
\end{barticle}
\endbibitem

\bibitem[\protect\citeauthoryear{Zama and Schwenker}{2023}]{zama2023ecg}
\begin{barticle}
\bauthor{\bsnm{Zama}, \binits{M.H.}},
\bauthor{\bsnm{Schwenker}, \binits{F.}}:
\batitle{Ecg synthesis via diffusion-based state space augmented transformer}.
\bjtitle{Sensors}
\bvolume{23}(\bissue{19}),
\bfpage{8328}
(\byear{2023})
\end{barticle}
\endbibitem

\bibitem[\protect\citeauthoryear{Wagner et~al.}{2020}]{wagner2020ptb}
\begin{barticle}
\bauthor{\bsnm{Wagner}, \binits{P.}},
\bauthor{\bsnm{Strodthoff}, \binits{N.}},
\bauthor{\bsnm{Bousseljot}, \binits{R.-D.}},
\bauthor{\bsnm{Kreiseler}, \binits{D.}},
\bauthor{\bsnm{Lunze}, \binits{F.I.}},
\bauthor{\bsnm{Samek}, \binits{W.}},
\bauthor{\bsnm{Schaeffter}, \binits{T.}}:
\batitle{Ptb-xl, a large publicly available electrocardiography dataset}.
\bjtitle{Scientific data}
\bvolume{7}(\bissue{1}),
\bfpage{1}--\blpage{15}
(\byear{2020})
\end{barticle}
\endbibitem

\bibitem[\protect\citeauthoryear{Shome et~al.}{2024}]{shome2024region}
\begin{bchapter}
\bauthor{\bsnm{Shome}, \binits{D.}},
\bauthor{\bsnm{Sarkar}, \binits{P.}},
\bauthor{\bsnm{Etemad}, \binits{A.}}:
\bctitle{Region-disentangled diffusion model for high-fidelity ppg-to-ecg translation}.
In: \bbtitle{Proceedings of the AAAI Conference on Artificial Intelligence},
vol. \bseriesno{38},
pp. \bfpage{15009}--\blpage{15019}
(\byear{2024})
\end{bchapter}
\endbibitem

\bibitem[\protect\citeauthoryear{Li et~al.}{2023}]{li2023descod}
\begin{botherref}
\oauthor{\bsnm{Li}, \binits{H.}},
\oauthor{\bsnm{Ditzler}, \binits{G.}},
\oauthor{\bsnm{Roveda}, \binits{J.}},
\oauthor{\bsnm{Li}, \binits{A.}}:
Descod-ecg: Deep score-based diffusion model for ecg baseline wander and noise removal.
IEEE Journal of Biomedical and Health Informatics
(2023)
\end{botherref}
\endbibitem

\bibitem[\protect\citeauthoryear{Akbari et~al.}{2021}]{akbari2021meta}
\begin{barticle}
\bauthor{\bsnm{Akbari}, \binits{A.}},
\bauthor{\bsnm{Martinez}, \binits{J.}},
\bauthor{\bsnm{Jafari}, \binits{R.}}:
\batitle{A meta-learning approach for fast personalization of modality translation models in wearable physiological sensing}.
\bjtitle{IEEE journal of biomedical and health informatics}
\bvolume{26}(\bissue{4}),
\bfpage{1516}--\blpage{1527}
(\byear{2021})
\end{barticle}
\endbibitem

\bibitem[\protect\citeauthoryear{Jia et~al.}{2022}]{jia2022personalized}
\begin{barticle}
\bauthor{\bsnm{Jia}, \binits{Z.}},
\bauthor{\bsnm{Shi}, \binits{Y.}},
\bauthor{\bsnm{Hu}, \binits{J.}}:
\batitle{Personalized neural network for patient-specific health monitoring in iot: a metalearning approach}.
\bjtitle{IEEE Transactions on Computer-Aided Design of Integrated Circuits and Systems}
\bvolume{41}(\bissue{12}),
\bfpage{5394}--\blpage{5407}
(\byear{2022})
\end{barticle}
\endbibitem

\bibitem[\protect\citeauthoryear{Liu et~al.}{2023}]{liu2023diagnosis}
\begin{barticle}
\bauthor{\bsnm{Liu}, \binits{Z.}},
\bauthor{\bsnm{Chen}, \binits{Y.}},
\bauthor{\bsnm{Zhang}, \binits{Y.}},
\bauthor{\bsnm{Ran}, \binits{S.}},
\bauthor{\bsnm{Cheng}, \binits{C.}},
\bauthor{\bsnm{Yang}, \binits{G.}}:
\batitle{Diagnosis of arrhythmias with few abnormal ecg samples using metric-based meta learning}.
\bjtitle{Computers in Biology and Medicine}
\bvolume{153},
\bfpage{106465}
(\byear{2023})
\end{barticle}
\endbibitem

\bibitem[\protect\citeauthoryear{Meqdad et~al.}{2022}]{meqdad2022meta}
\begin{barticle}
\bauthor{\bsnm{Meqdad}, \binits{M.N.}},
\bauthor{\bsnm{Abdali-Mohammadi}, \binits{F.}},
\bauthor{\bsnm{Kadry}, \binits{S.}}:
\batitle{Meta structural learning algorithm with interpretable convolutional neural networks for arrhythmia detection of multisession ecg}.
\bjtitle{IEEE Access}
\bvolume{10},
\bfpage{61410}--\blpage{61425}
(\byear{2022})
\end{barticle}
\endbibitem

\bibitem[\protect\citeauthoryear{Musa et~al.}{2023}]{musa2023systematic}
\begin{barticle}
\bauthor{\bsnm{Musa}, \binits{N.}},
\bauthor{\bsnm{Gital}, \binits{A.Y.}},
\bauthor{\bsnm{Aljojo}, \binits{N.}},
\bauthor{\bsnm{Chiroma}, \binits{H.}},
\bauthor{\bsnm{Adewole}, \binits{K.S.}},
\bauthor{\bsnm{Mojeed}, \binits{H.A.}},
\bauthor{\bsnm{Faruk}, \binits{N.}},
\bauthor{\bsnm{Abdulkarim}, \binits{A.}},
\bauthor{\bsnm{Emmanuel}, \binits{I.}},
\bauthor{\bsnm{Folawiyo}, \binits{Y.Y.}}, \betal:
\batitle{A systematic review and meta-data analysis on the applications of deep learning in electrocardiogram}.
\bjtitle{Journal of ambient intelligence and humanized computing}
\bvolume{14}(\bissue{7}),
\bfpage{9677}--\blpage{9750}
(\byear{2023})
\end{barticle}
\endbibitem

\bibitem[\protect\citeauthoryear{Sun et~al.}{2023}]{sun2023federated}
\begin{barticle}
\bauthor{\bsnm{Sun}, \binits{L.}},
\bauthor{\bsnm{Wu}, \binits{J.}},
\bauthor{\bsnm{Xu}, \binits{Y.}},
\bauthor{\bsnm{Zhang}, \binits{Y.}}:
\batitle{A federated learning and blockchain framework for physiological signal classification based on continual learning}.
\bjtitle{Information Sciences}
\bvolume{630},
\bfpage{586}--\blpage{598}
(\byear{2023})
\end{barticle}
\endbibitem

\bibitem[\protect\citeauthoryear{Liu et~al.}{2021}]{liu2021few}
\begin{barticle}
\bauthor{\bsnm{Liu}, \binits{T.}},
\bauthor{\bsnm{Yang}, \binits{Y.}},
\bauthor{\bsnm{Fan}, \binits{W.}},
\bauthor{\bsnm{Wu}, \binits{C.}}:
\batitle{Few-shot learning for cardiac arrhythmia detection based on electrocardiogram data from wearable devices}.
\bjtitle{Digital Signal Processing}
\bvolume{116},
\bfpage{103094}
(\byear{2021})
\end{barticle}
\endbibitem

\bibitem[\protect\citeauthoryear{Rjoob et~al.}{2022}]{rjoob2022machine}
\begin{barticle}
\bauthor{\bsnm{Rjoob}, \binits{K.}},
\bauthor{\bsnm{Bond}, \binits{R.}},
\bauthor{\bsnm{Finlay}, \binits{D.}},
\bauthor{\bsnm{McGilligan}, \binits{V.}},
\bauthor{\bsnm{Leslie}, \binits{S.J.}},
\bauthor{\bsnm{Rababah}, \binits{A.}},
\bauthor{\bsnm{Iftikhar}, \binits{A.}},
\bauthor{\bsnm{Guldenring}, \binits{D.}},
\bauthor{\bsnm{Knoery}, \binits{C.}},
\bauthor{\bsnm{McShane}, \binits{A.}}, \betal:
\batitle{Machine learning and the electrocardiogram over two decades: Time series and meta-analysis of the algorithms, evaluation metrics and applications}.
\bjtitle{Artificial Intelligence in Medicine}
\bvolume{132},
\bfpage{102381}
(\byear{2022})
\end{barticle}
\endbibitem

\bibitem[\protect\citeauthoryear{Sun et~al.}{2023}]{sun2023few}
\begin{botherref}
\oauthor{\bsnm{Sun}, \binits{L.}},
\oauthor{\bsnm{Zhang}, \binits{M.}},
\oauthor{\bsnm{Wang}, \binits{B.}},
\oauthor{\bsnm{Tiwari}, \binits{P.}}:
Few-shot class-incremental learning for medical time series classification.
IEEE journal of biomedical and health informatics
(2023)
\end{botherref}
\endbibitem

\bibitem[\protect\citeauthoryear{Essa and Xie}{2021}]{essa2021ensemble}
\begin{barticle}
\bauthor{\bsnm{Essa}, \binits{E.}},
\bauthor{\bsnm{Xie}, \binits{X.}}:
\batitle{An ensemble of deep learning-based multi-model for ecg heartbeats arrhythmia classification}.
\bjtitle{ieee access}
\bvolume{9},
\bfpage{103452}--\blpage{103464}
(\byear{2021})
\end{barticle}
\endbibitem

\bibitem[\protect\citeauthoryear{Hoi et~al.}{2021}]{hoi2021online}
\begin{barticle}
\bauthor{\bsnm{Hoi}, \binits{S.C.}},
\bauthor{\bsnm{Sahoo}, \binits{D.}},
\bauthor{\bsnm{Lu}, \binits{J.}},
\bauthor{\bsnm{Zhao}, \binits{P.}}:
\batitle{Online learning: A comprehensive survey}.
\bjtitle{Neurocomputing}
\bvolume{459},
\bfpage{249}--\blpage{289}
(\byear{2021})
\end{barticle}
\endbibitem

\bibitem[\protect\citeauthoryear{Bonomi et~al.}{2022}]{bonomi2022sharing}
\begin{barticle}
\bauthor{\bsnm{Bonomi}, \binits{L.}},
\bauthor{\bsnm{Wu}, \binits{Z.}},
\bauthor{\bsnm{Fan}, \binits{L.}}:
\batitle{Sharing personal ecg time-series data privately}.
\bjtitle{Journal of the American Medical Informatics Association}
\bvolume{29}(\bissue{7}),
\bfpage{1152}--\blpage{1160}
(\byear{2022})
\end{barticle}
\endbibitem

\bibitem[\protect\citeauthoryear{Ngu et~al.}{2021}]{ngu2021iot}
\begin{barticle}
\bauthor{\bsnm{Ngu}, \binits{A.H.}},
\bauthor{\bsnm{Eyitayo}, \binits{J.S.}},
\bauthor{\bsnm{Yang}, \binits{G.}},
\bauthor{\bsnm{Campbell}, \binits{C.}},
\bauthor{\bsnm{Sheng}, \binits{Q.Z.}},
\bauthor{\bsnm{Ni}, \binits{J.}}:
\batitle{An iot edge computing framework using cordova accessor host}.
\bjtitle{IEEE Internet of Things Journal}
\bvolume{9}(\bissue{1}),
\bfpage{671}--\blpage{683}
(\byear{2021})
\end{barticle}
\endbibitem

\bibitem[\protect\citeauthoryear{Ni et~al.}{2022a}]{ni2022cross}
\begin{bchapter}
\bauthor{\bsnm{Ni}, \binits{J.}},
\bauthor{\bsnm{Sarbajna}, \binits{R.}},
\bauthor{\bsnm{Liu}, \binits{Y.}},
\bauthor{\bsnm{Ngu}, \binits{A.H.}},
\bauthor{\bsnm{Yan}, \binits{Y.}}:
\bctitle{Cross-modal knowledge distillation for vision-to-sensor action recognition}.
In: \bbtitle{ICASSP 2022-2022 IEEE International Conference on Acoustics, Speech and Signal Processing (ICASSP)},
pp. \bfpage{4448}--\blpage{4452}
(\byear{2022}).
\bcomment{IEEE}
\end{bchapter}
\endbibitem

\bibitem[\protect\citeauthoryear{Ni et~al.}{2022b}]{ni2022progressive}
\begin{bchapter}
\bauthor{\bsnm{Ni}, \binits{J.}},
\bauthor{\bsnm{Ngu}, \binits{A.H.}},
\bauthor{\bsnm{Yan}, \binits{Y.}}:
\bctitle{Progressive cross-modal knowledge distillation for human action recognition}.
In: \bbtitle{Proceedings of the 30th ACM International Conference on Multimedia},
pp. \bfpage{5903}--\blpage{5912}
(\byear{2022})
\end{bchapter}
\endbibitem

\bibitem[\protect\citeauthoryear{Mehari and Strodthoff}{2023}]{mehari2023towards}
\begin{botherref}
\oauthor{\bsnm{Mehari}, \binits{T.}},
\oauthor{\bsnm{Strodthoff}, \binits{N.}}:
Towards quantitative precision for ecg analysis: Leveraging state space models, self-supervision and patient metadata.
IEEE Journal of Biomedical and Health Informatics
(2023)
\end{botherref}
\endbibitem

\bibitem[\protect\citeauthoryear{Patro and Agneeswaran}{2024}]{patro2024mamba}
\begin{botherref}
\oauthor{\bsnm{Patro}, \binits{B.N.}},
\oauthor{\bsnm{Agneeswaran}, \binits{V.S.}}:
Mamba-360: Survey of state space models as transformer alternative for long sequence modelling: Methods, applications, and challenges.
arXiv preprint arXiv:2404.16112
(2024)
\end{botherref}
\endbibitem

\bibitem[\protect\citeauthoryear{Ni et~al.}{2024}]{ni2024adaptive}
\begin{bchapter}
\bauthor{\bsnm{Ni}, \binits{J.}},
\bauthor{\bsnm{Tang}, \binits{H.}},
\bauthor{\bsnm{Shang}, \binits{Y.}},
\bauthor{\bsnm{Duan}, \binits{B.}},
\bauthor{\bsnm{Yan}, \binits{Y.}}:
\bctitle{Adaptive cross-architecture mutual knowledge distillation}.
In: \bbtitle{2024 IEEE 18th International Conference on Automatic Face and Gesture Recognition (FG)},
pp. \bfpage{1}--\blpage{5}
(\byear{2024}).
\bcomment{IEEE}
\end{bchapter}
\endbibitem

\bibitem[\protect\citeauthoryear{Hong et~al.}{2020}]{hong2020opportunities}
\begin{barticle}
\bauthor{\bsnm{Hong}, \binits{S.}},
\bauthor{\bsnm{Zhou}, \binits{Y.}},
\bauthor{\bsnm{Shang}, \binits{J.}},
\bauthor{\bsnm{Xiao}, \binits{C.}},
\bauthor{\bsnm{Sun}, \binits{J.}}:
\batitle{Opportunities and challenges of deep learning methods for electrocardiogram data: A systematic review}.
\bjtitle{Computers in biology and medicine}
\bvolume{122},
\bfpage{103801}
(\byear{2020})
\end{barticle}
\endbibitem

\bibitem[\protect\citeauthoryear{Thambawita et~al.}{2021}]{thambawita2021deepfake}
\begin{barticle}
\bauthor{\bsnm{Thambawita}, \binits{V.}},
\bauthor{\bsnm{Isaksen}, \binits{J.L.}},
\bauthor{\bsnm{Hicks}, \binits{S.A.}},
\bauthor{\bsnm{Ghouse}, \binits{J.}},
\bauthor{\bsnm{Ahlberg}, \binits{G.}},
\bauthor{\bsnm{Linneberg}, \binits{A.}},
\bauthor{\bsnm{Grarup}, \binits{N.}},
\bauthor{\bsnm{Ellervik}, \binits{C.}},
\bauthor{\bsnm{Olesen}, \binits{M.S.}},
\bauthor{\bsnm{Hansen}, \binits{T.}}, \betal:
\batitle{Deepfake electrocardiograms using generative adversarial networks are the beginning of the end for privacy issues in medicine}.
\bjtitle{Scientific reports}
\bvolume{11}(\bissue{1}),
\bfpage{21896}
(\byear{2021})
\end{barticle}
\endbibitem

\bibitem[\protect\citeauthoryear{Mason et~al.}{2024}]{mason2024ai}
\begin{barticle}
\bauthor{\bsnm{Mason}, \binits{F.}},
\bauthor{\bsnm{Pandey}, \binits{A.C.}},
\bauthor{\bsnm{Gadaleta}, \binits{M.}},
\bauthor{\bsnm{Topol}, \binits{E.J.}},
\bauthor{\bsnm{Muse}, \binits{E.D.}},
\bauthor{\bsnm{Quer}, \binits{G.}}:
\batitle{Ai-enhanced reconstruction of the 12-lead electrocardiogram via 3-leads with accurate clinical assessment}.
\bjtitle{npj Digital Medicine}
\bvolume{7}(\bissue{1}),
\bfpage{201}
(\byear{2024})
\end{barticle}
\endbibitem

\bibitem[\protect\citeauthoryear{Alahmadi et~al.}{2021}]{alahmadi2021explainable}
\begin{barticle}
\bauthor{\bsnm{Alahmadi}, \binits{A.}},
\bauthor{\bsnm{Davies}, \binits{A.}},
\bauthor{\bsnm{Royle}, \binits{J.}},
\bauthor{\bsnm{Goodwin}, \binits{L.}},
\bauthor{\bsnm{Cresswell}, \binits{K.}},
\bauthor{\bsnm{Arain}, \binits{Z.}},
\bauthor{\bsnm{Vigo}, \binits{M.}},
\bauthor{\bsnm{Jay}, \binits{C.}}:
\batitle{An explainable algorithm for detecting drug-induced qt-prolongation at risk of torsades de pointes (tdp) regardless of heart rate and t-wave morphology}.
\bjtitle{Computers in Biology and Medicine}
\bvolume{131},
\bfpage{104281}
(\byear{2021})
\end{barticle}
\endbibitem

\bibitem[\protect\citeauthoryear{Ma et~al.}{2024}]{ma2024evolution}
\begin{botherref}
\oauthor{\bsnm{Ma}, \binits{W.}},
\oauthor{\bsnm{Sheng}, \binits{B.}},
\oauthor{\bsnm{Liu}, \binits{Y.}},
\oauthor{\bsnm{Qian}, \binits{J.}},
\oauthor{\bsnm{Liu}, \binits{X.}},
\oauthor{\bsnm{Li}, \binits{J.}},
\oauthor{\bsnm{Ouyang}, \binits{D.}},
\oauthor{\bsnm{Wang}, \binits{H.}},
\oauthor{\bsnm{Atanasov}, \binits{A.G.}},
\oauthor{\bsnm{Keane}, \binits{P.A.}}, et al.:
Evolution of Future Medical AI Models—From Task-Specific, Disease-Centric to Universal Health.
Massachusetts Medical Society
(2024)
\end{botherref}
\endbibitem

\bibitem[\protect\citeauthoryear{Topol}{2019}]{topol2019high}
\begin{barticle}
\bauthor{\bsnm{Topol}, \binits{E.J.}}:
\batitle{High-performance medicine: the convergence of human and artificial intelligence}.
\bjtitle{Nature medicine}
\bvolume{25}(\bissue{1}),
\bfpage{44}--\blpage{56}
(\byear{2019})
\end{barticle}
\endbibitem

\bibitem[\protect\citeauthoryear{Dai et~al.}{2024}]{dai2024deep}
\begin{barticle}
\bauthor{\bsnm{Dai}, \binits{L.}},
\bauthor{\bsnm{Sheng}, \binits{B.}},
\bauthor{\bsnm{Chen}, \binits{T.}},
\bauthor{\bsnm{Wu}, \binits{Q.}},
\bauthor{\bsnm{Liu}, \binits{R.}},
\bauthor{\bsnm{Cai}, \binits{C.}},
\bauthor{\bsnm{Wu}, \binits{L.}},
\bauthor{\bsnm{Yang}, \binits{D.}},
\bauthor{\bsnm{Hamzah}, \binits{H.}},
\bauthor{\bsnm{Liu}, \binits{Y.}}, \betal:
\batitle{A deep learning system for predicting time to progression of diabetic retinopathy}.
\bjtitle{Nature Medicine}
\bvolume{30}(\bissue{2}),
\bfpage{584}--\blpage{594}
(\byear{2024})
\end{barticle}
\endbibitem

\bibitem[\protect\citeauthoryear{Li et~al.}{2024}]{li2024integrated}
\begin{botherref}
\oauthor{\bsnm{Li}, \binits{J.}},
\oauthor{\bsnm{Guan}, \binits{Z.}},
\oauthor{\bsnm{Wang}, \binits{J.}},
\oauthor{\bsnm{Cheung}, \binits{C.Y.}},
\oauthor{\bsnm{Zheng}, \binits{Y.}},
\oauthor{\bsnm{Lim}, \binits{L.-L.}},
\oauthor{\bsnm{Lim}, \binits{C.C.}},
\oauthor{\bsnm{Ruamviboonsuk}, \binits{P.}},
\oauthor{\bsnm{Raman}, \binits{R.}},
\oauthor{\bsnm{Corsino}, \binits{L.}}, et al.:
Integrated image-based deep learning and language models for primary diabetes care.
Nature Medicine,
1--11
(2024)
\end{botherref}
\endbibitem

\bibitem[\protect\citeauthoryear{Khera}{2024}]{khera2024ai}
\begin{botherref}
\oauthor{\bsnm{Khera}, \binits{R.}}:
AI-enabled diagnosis from an electrocardiogram image: the next frontier of innovation in a century-old technology.
BMJ Publishing Group Ltd and British Cardiovascular Society
(2024)
\end{botherref}
\endbibitem

\bibitem[\protect\citeauthoryear{Rajpurkar et~al.}{2022}]{rajpurkar2022ai}
\begin{barticle}
\bauthor{\bsnm{Rajpurkar}, \binits{P.}},
\bauthor{\bsnm{Chen}, \binits{E.}},
\bauthor{\bsnm{Banerjee}, \binits{O.}},
\bauthor{\bsnm{Topol}, \binits{E.J.}}:
\batitle{Ai in health and medicine}.
\bjtitle{Nature medicine}
\bvolume{28}(\bissue{1}),
\bfpage{31}--\blpage{38}
(\byear{2022})
\end{barticle}
\endbibitem

\bibitem[\protect\citeauthoryear{Jia et~al.}{2020}]{jia2020personalized}
\begin{bchapter}
\bauthor{\bsnm{Jia}, \binits{Z.}},
\bauthor{\bsnm{Wang}, \binits{Z.}},
\bauthor{\bsnm{Hong}, \binits{F.}},
\bauthor{\bsnm{Ping}, \binits{L.}},
\bauthor{\bsnm{Shi}, \binits{Y.}},
\bauthor{\bsnm{Hu}, \binits{J.}}:
\bctitle{Personalized deep learning for ventricular arrhythmias detection on medical iot systems}.
In: \bbtitle{Proceedings of the 39th International Conference on Computer-aided Design},
pp. \bfpage{1}--\blpage{9}
(\byear{2020})
\end{bchapter}
\endbibitem

\bibitem[\protect\citeauthoryear{Jia et~al.}{2021}]{jia2021enabling}
\begin{bchapter}
\bauthor{\bsnm{Jia}, \binits{Z.}},
\bauthor{\bsnm{Hong}, \binits{F.}},
\bauthor{\bsnm{Ping}, \binits{L.}},
\bauthor{\bsnm{Shi}, \binits{Y.}},
\bauthor{\bsnm{Hu}, \binits{J.}}:
\bctitle{Enabling on-device model personalization for ventricular arrhythmias detection by generative adversarial networks}.
In: \bbtitle{2021 58th ACM/IEEE Design Automation Conference (DAC)},
pp. \bfpage{163}--\blpage{168}
(\byear{2021}).
\bcomment{IEEE}
\end{bchapter}
\endbibitem

\bibitem[\protect\citeauthoryear{Qin et~al.}{2024}]{qin2024lightweight}
\begin{barticle}
\bauthor{\bsnm{Qin}, \binits{K.}},
\bauthor{\bsnm{Huang}, \binits{W.}},
\bauthor{\bsnm{Zhang}, \binits{T.}},
\bauthor{\bsnm{Zhang}, \binits{H.}},
\bauthor{\bsnm{Cheng}, \binits{X.}}:
\batitle{A lightweight selfonn model for general ecg classification with pretraining}.
\bjtitle{Biomedical Signal Processing and Control}
\bvolume{89},
\bfpage{105780}
(\byear{2024})
\end{barticle}
\endbibitem

\bibitem[\protect\citeauthoryear{Sun et~al.}{2022}]{sun2022perae}
\begin{barticle}
\bauthor{\bsnm{Sun}, \binits{L.}},
\bauthor{\bsnm{Zhong}, \binits{Z.}},
\bauthor{\bsnm{Qu}, \binits{Z.}},
\bauthor{\bsnm{Xiong}, \binits{N.}}:
\batitle{Perae: an effective personalized autoencoder for ecg-based biometric in augmented reality system}.
\bjtitle{IEEE journal of biomedical and health informatics}
\bvolume{26}(\bissue{6}),
\bfpage{2435}--\blpage{2446}
(\byear{2022})
\end{barticle}
\endbibitem

\bibitem[\protect\citeauthoryear{Xu et~al.}{2022}]{xu2022inter}
\begin{barticle}
\bauthor{\bsnm{Xu}, \binits{S.S.}},
\bauthor{\bsnm{Mak}, \binits{M.-W.}},
\bauthor{\bsnm{Chang}, \binits{C.}}:
\batitle{Inter-patient ecg classification with i-vector based unsupervised patient adaptation}.
\bjtitle{Expert Systems with Applications}
\bvolume{210},
\bfpage{118410}
(\byear{2022})
\end{barticle}
\endbibitem

\bibitem[\protect\citeauthoryear{Chao et~al.}{2023}]{chao2023research}
\begin{bchapter}
\bauthor{\bsnm{Chao}, \binits{L.}},
\bauthor{\bsnm{Li}, \binits{Z.}},
\bauthor{\bsnm{Zhang}, \binits{H.}}:
\bctitle{Research on arrhythmia classification based on domain adaptation}.
In: \bbtitle{2023 9th International Conference on Control, Automation and Robotics (ICCAR)},
pp. \bfpage{147}--\blpage{152}
(\byear{2023}).
\bcomment{IEEE}
\end{bchapter}
\endbibitem

\bibitem[\protect\citeauthoryear{Tian et~al.}{2022}]{tian2022cross}
\begin{barticle}
\bauthor{\bsnm{Tian}, \binits{X.}},
\bauthor{\bsnm{Zhu}, \binits{Q.}},
\bauthor{\bsnm{Li}, \binits{Y.}},
\bauthor{\bsnm{Wu}, \binits{M.}}:
\batitle{Cross-domain joint dictionary learning for ecg inference from ppg}.
\bjtitle{IEEE Internet of Things Journal}
\bvolume{10}(\bissue{9}),
\bfpage{8140}--\blpage{8154}
(\byear{2022})
\end{barticle}
\endbibitem

\bibitem[\protect\citeauthoryear{Zhai et~al.}{2020}]{zhai2020semi}
\begin{barticle}
\bauthor{\bsnm{Zhai}, \binits{X.}},
\bauthor{\bsnm{Zhou}, \binits{Z.}},
\bauthor{\bsnm{Tin}, \binits{C.}}:
\batitle{Semi-supervised learning for ecg classification without patient-specific labeled data}.
\bjtitle{Expert Systems with Applications}
\bvolume{158},
\bfpage{113411}
(\byear{2020})
\end{barticle}
\endbibitem

\bibitem[\protect\citeauthoryear{Wu et~al.}{2023}]{wu2023udama}
\begin{bchapter}
\bauthor{\bsnm{Wu}, \binits{Y.}},
\bauthor{\bsnm{Spathis}, \binits{D.}},
\bauthor{\bsnm{Jia}, \binits{H.}},
\bauthor{\bsnm{Perez-Pozuelo}, \binits{I.}},
\bauthor{\bsnm{Gonzales}, \binits{T.I.}},
\bauthor{\bsnm{Brage}, \binits{S.}},
\bauthor{\bsnm{Wareham}, \binits{N.}},
\bauthor{\bsnm{Mascolo}, \binits{C.}}:
\bctitle{Udama: Unsupervised domain adaptation through multi-discriminator adversarial training with noisy labels improves cardio-fitness prediction}.
In: \bbtitle{Machine Learning for Healthcare Conference},
pp. \bfpage{863}--\blpage{883}
(\byear{2023}).
\bcomment{PMLR}
\end{bchapter}
\endbibitem

\bibitem[\protect\citeauthoryear{De~Giovanni et~al.}{2022}]{de2022adaptive}
\begin{barticle}
\bauthor{\bsnm{De~Giovanni}, \binits{E.}},
\bauthor{\bsnm{Teijeiro}, \binits{T.}},
\bauthor{\bsnm{Millet}, \binits{G.P.}},
\bauthor{\bsnm{Atienza}, \binits{D.}}:
\batitle{Adaptive r-peak detection on wearable ecg sensors for high-intensity exercise}.
\bjtitle{IEEE Transactions on Biomedical Engineering}
\bvolume{70}(\bissue{3}),
\bfpage{941}--\blpage{953}
(\byear{2022})
\end{barticle}
\endbibitem

\bibitem[\protect\citeauthoryear{Kantharaju et~al.}{2023}]{kantharaju2023framework}
\begin{botherref}
\oauthor{\bsnm{Kantharaju}, \binits{P.}},
\oauthor{\bsnm{Vakacherla}, \binits{S.S.}},
\oauthor{\bsnm{Jacobson}, \binits{M.}},
\oauthor{\bsnm{Jeong}, \binits{H.}},
\oauthor{\bsnm{Mevada}, \binits{M.N.}},
\oauthor{\bsnm{Zhou}, \binits{X.}},
\oauthor{\bsnm{Major}, \binits{M.J.}},
\oauthor{\bsnm{Kim}, \binits{M.}}:
Framework for personalizing wearable devices using real-time physiological measures.
IEEE Access
(2023)
\end{botherref}
\endbibitem

\bibitem[\protect\citeauthoryear{Scrugli et~al.}{2021}]{scrugli2021adaptive}
\begin{barticle}
\bauthor{\bsnm{Scrugli}, \binits{M.A.}},
\bauthor{\bsnm{Loi}, \binits{D.}},
\bauthor{\bsnm{Raffo}, \binits{L.}},
\bauthor{\bsnm{Meloni}, \binits{P.}}:
\batitle{An adaptive cognitive sensor node for ecg monitoring in the internet of medical things}.
\bjtitle{IEEE Access}
\bvolume{10},
\bfpage{1688}--\blpage{1705}
(\byear{2021})
\end{barticle}
\endbibitem

\bibitem[\protect\citeauthoryear{Li et~al.}{2021}]{li2021mixup}
\begin{barticle}
\bauthor{\bsnm{Li}, \binits{J.}},
\bauthor{\bsnm{Wang}, \binits{G.}},
\bauthor{\bsnm{Chen}, \binits{M.}},
\bauthor{\bsnm{Ding}, \binits{Z.}},
\bauthor{\bsnm{Yang}, \binits{H.}}:
\batitle{Mixup asymmetric tri-training for heartbeat classification under domain shift}.
\bjtitle{IEEE Signal Processing Letters}
\bvolume{28},
\bfpage{718}--\blpage{722}
(\byear{2021})
\end{barticle}
\endbibitem

\bibitem[\protect\citeauthoryear{Wang et~al.}{2023}]{wang2023adversarial}
\begin{botherref}
\oauthor{\bsnm{Wang}, \binits{N.}},
\oauthor{\bsnm{Feng}, \binits{P.}},
\oauthor{\bsnm{Ge}, \binits{Z.}},
\oauthor{\bsnm{Zhou}, \binits{Y.}},
\oauthor{\bsnm{Zhou}, \binits{B.}},
\oauthor{\bsnm{Wang}, \binits{Z.}}:
Adversarial spatiotemporal contrastive learning for electrocardiogram signals.
IEEE Transactions on Neural Networks and Learning Systems
(2023)
\end{botherref}
\endbibitem

\bibitem[\protect\citeauthoryear{Kumari et~al.}{2022}]{kumari2022classification}
\begin{barticle}
\bauthor{\bsnm{Kumari}, \binits{L.}},
\bauthor{\bsnm{Sai}, \binits{Y.P.}}, \betal:
\batitle{Classification of ecg beats using optimized decision tree and adaptive boosted optimized decision tree}.
\bjtitle{Signal, Image and Video Processing}
\bvolume{16}(\bissue{3}),
\bfpage{695}--\blpage{703}
(\byear{2022})
\end{barticle}
\endbibitem

\bibitem[\protect\citeauthoryear{Rajani~Kumari and Chalapathi~Rao}{2023}]{rajani2023ecg}
\begin{barticle}
\bauthor{\bsnm{Rajani~Kumari}, \binits{L.}},
\bauthor{\bsnm{Chalapathi~Rao}, \binits{Y.}}:
\batitle{Ecg beat classification using proposed pattern adaptive wavelet-based hybrid classifiers}.
\bjtitle{Signal, Image and Video Processing}
\bvolume{17}(\bissue{6}),
\bfpage{2827}--\blpage{2835}
(\byear{2023})
\end{barticle}
\endbibitem

\bibitem[\protect\citeauthoryear{Xia et~al.}{2023}]{xia2023generative}
\begin{barticle}
\bauthor{\bsnm{Xia}, \binits{Y.}},
\bauthor{\bsnm{Xu}, \binits{Y.}},
\bauthor{\bsnm{Chen}, \binits{P.}},
\bauthor{\bsnm{Zhang}, \binits{J.}},
\bauthor{\bsnm{Zhang}, \binits{Y.}}:
\batitle{Generative adversarial network with transformer generator for boosting ecg classification}.
\bjtitle{Biomedical Signal Processing and Control}
\bvolume{80},
\bfpage{104276}
(\byear{2023})
\end{barticle}
\endbibitem

\bibitem[\protect\citeauthoryear{Dasgupta et~al.}{2021}]{dasgupta2021cardiogan}
\begin{bchapter}
\bauthor{\bsnm{Dasgupta}, \binits{S.}},
\bauthor{\bsnm{Das}, \binits{S.}},
\bauthor{\bsnm{Bhattacharya}, \binits{U.}}:
\bctitle{Cardiogan: an attention-based generative adversarial network for generation of electrocardiograms}.
In: \bbtitle{2020 25th International Conference on Pattern Recognition (ICPR)},
pp. \bfpage{3193}--\blpage{3200}
(\byear{2021}).
\bcomment{IEEE}
\end{bchapter}
\endbibitem

\bibitem[\protect\citeauthoryear{Yoon and Joo}{2024}]{yoon2024classification}
\begin{barticle}
\bauthor{\bsnm{Yoon}, \binits{G.-W.}},
\bauthor{\bsnm{Joo}, \binits{S.}}:
\batitle{Classification feasibility test on multi-lead electrocardiography signals generated from single-lead electrocardiography signals}.
\bjtitle{Scientific Reports}
\bvolume{14}(\bissue{1}),
\bfpage{1888}
(\byear{2024})
\end{barticle}
\endbibitem

\bibitem[\protect\citeauthoryear{Qin et~al.}{2023}]{qin2023novel}
\begin{barticle}
\bauthor{\bsnm{Qin}, \binits{J.}},
\bauthor{\bsnm{Gao}, \binits{F.}},
\bauthor{\bsnm{Wang}, \binits{Z.}},
\bauthor{\bsnm{Wong}, \binits{D.C.}},
\bauthor{\bsnm{Zhao}, \binits{Z.}},
\bauthor{\bsnm{Relton}, \binits{S.D.}},
\bauthor{\bsnm{Fang}, \binits{H.}}:
\batitle{A novel temporal generative adversarial network for electrocardiography anomaly detection}.
\bjtitle{Artificial Intelligence in Medicine}
\bvolume{136},
\bfpage{102489}
(\byear{2023})
\end{barticle}
\endbibitem

\bibitem[\protect\citeauthoryear{Simone and Bacciu}{2023}]{simone2023ecgan}
\begin{bchapter}
\bauthor{\bsnm{Simone}, \binits{L.}},
\bauthor{\bsnm{Bacciu}, \binits{D.}}:
\bctitle{Ecgan: Self-supervised generative adversarial network for electrocardiography}.
In: \bbtitle{International Conference on Artificial Intelligence in Medicine},
pp. \bfpage{276}--\blpage{280}
(\byear{2023}).
\bcomment{Springer}
\end{bchapter}
\endbibitem

\bibitem[\protect\citeauthoryear{Monachino et~al.}{2023}]{monachino2023deep}
\begin{botherref}
\oauthor{\bsnm{Monachino}, \binits{G.}},
\oauthor{\bsnm{Zanchi}, \binits{B.}},
\oauthor{\bsnm{Fiorillo}, \binits{L.}},
\oauthor{\bsnm{Conte}, \binits{G.}},
\oauthor{\bsnm{Auricchio}, \binits{A.}},
\oauthor{\bsnm{Tzovara}, \binits{A.}},
\oauthor{\bsnm{Faraci}, \binits{F.D.}}:
Deep generative models: The winning key for large and easily accessible ecg datasets?
Computers in biology and medicine,
107655
(2023)
\end{botherref}
\endbibitem

\bibitem[\protect\citeauthoryear{Cao et~al.}{2020}]{cao2020feature}
\begin{bchapter}
\bauthor{\bsnm{Cao}, \binits{F.}},
\bauthor{\bsnm{Budhota}, \binits{A.}},
\bauthor{\bsnm{Chen}, \binits{H.}},
\bauthor{\bsnm{Rajput}, \binits{K.S.}}:
\bctitle{Feature matching based ecg generative network for arrhythmia event augmentation}.
In: \bbtitle{2020 42nd Annual International Conference of the IEEE Engineering in Medicine \& Biology Society (EMBC)},
pp. \bfpage{296}--\blpage{299}
(\byear{2020}).
\bcomment{IEEE}
\end{bchapter}
\endbibitem

\bibitem[\protect\citeauthoryear{Sun et~al.}{2021}]{sun2021beatclass}
\begin{barticle}
\bauthor{\bsnm{Sun}, \binits{L.}},
\bauthor{\bsnm{Wang}, \binits{Y.}},
\bauthor{\bsnm{Qu}, \binits{Z.}},
\bauthor{\bsnm{Xiong}, \binits{N.N.}}:
\batitle{Beatclass: a sustainable ecg classification system in iot-based ehealth}.
\bjtitle{IEEE Internet of Things Journal}
\bvolume{9}(\bissue{10}),
\bfpage{7178}--\blpage{7195}
(\byear{2021})
\end{barticle}
\endbibitem

\bibitem[\protect\citeauthoryear{Zhou et~al.}{2021}]{zhou2021fully}
\begin{barticle}
\bauthor{\bsnm{Zhou}, \binits{Z.}},
\bauthor{\bsnm{Zhai}, \binits{X.}},
\bauthor{\bsnm{Tin}, \binits{C.}}:
\batitle{Fully automatic electrocardiogram classification system based on generative adversarial network with auxiliary classifier}.
\bjtitle{Expert Systems with Applications}
\bvolume{174},
\bfpage{114809}
(\byear{2021})
\end{barticle}
\endbibitem

\bibitem[\protect\citeauthoryear{Sanamdikar et~al.}{2022}]{sanamdikar2022classification}
\begin{bchapter}
\bauthor{\bsnm{Sanamdikar}, \binits{S.}},
\bauthor{\bsnm{Karajanagi}, \binits{N.}},
\bauthor{\bsnm{Kowdiki}, \binits{K.}},
\bauthor{\bsnm{Kamble}, \binits{S.}}:
\bctitle{Classification of ecg signal for cardiac arrhythmia detection using gan method}.
In: \bbtitle{Intelligent Communication Technologies and Virtual Mobile Networks: Proceedings of ICICV 2022},
pp. \bfpage{257}--\blpage{271}
(\byear{2022}).
\bcomment{Springer}
\end{bchapter}
\endbibitem

\bibitem[\protect\citeauthoryear{Yoon and Joo}{2023}]{yoon2023generated}
\begin{bchapter}
\bauthor{\bsnm{Yoon}, \binits{G.-W.}},
\bauthor{\bsnm{Joo}, \binits{S.}}:
\bctitle{Generated ecg signal feasibility evaluation for classification}.
In: \bbtitle{2023 Computing in Cardiology (CinC)},
vol. \bseriesno{50},
pp. \bfpage{1}--\blpage{4}
(\byear{2023}).
\bcomment{IEEE}
\end{bchapter}
\endbibitem

\bibitem[\protect\citeauthoryear{Gyawali et~al.}{2021}]{gyawali2021learning}
\begin{barticle}
\bauthor{\bsnm{Gyawali}, \binits{P.K.}},
\bauthor{\bsnm{Murkute}, \binits{J.V.}},
\bauthor{\bsnm{Toloubidokhti}, \binits{M.}},
\bauthor{\bsnm{Jiang}, \binits{X.}},
\bauthor{\bsnm{Horacek}, \binits{B.M.}},
\bauthor{\bsnm{Sapp}, \binits{J.L.}},
\bauthor{\bsnm{Wang}, \binits{L.}}:
\batitle{Learning to disentangle inter-subject anatomical variations in electrocardiographic data}.
\bjtitle{IEEE Transactions on Biomedical Engineering}
\bvolume{69}(\bissue{2}),
\bfpage{860}--\blpage{870}
(\byear{2021})
\end{barticle}
\endbibitem

\bibitem[\protect\citeauthoryear{Nankani and Dutta~Baruah}{2022}]{nankani2022improved}
\begin{bchapter}
\bauthor{\bsnm{Nankani}, \binits{D.}},
\bauthor{\bsnm{Dutta~Baruah}, \binits{R.}}:
\bctitle{Improved diagnostic performance of arrhythmia classification using conditional gan augmented heartbeats}.
In: \bbtitle{Generative Adversarial Learning: Architectures and Applications},
pp. \bfpage{275}--\blpage{304}
(\byear{2022}).
\bcomment{Springer}
\end{bchapter}
\endbibitem

\bibitem[\protect\citeauthoryear{Agrawal et~al.}{2022}]{agrawal2022ecg}
\begin{barticle}
\bauthor{\bsnm{Agrawal}, \binits{A.}},
\bauthor{\bsnm{Chauhan}, \binits{A.}},
\bauthor{\bsnm{Shetty}, \binits{M.K.}},
\bauthor{\bsnm{Gupta}, \binits{M.D.}},
\bauthor{\bsnm{Gupta}, \binits{A.}}, \betal:
\batitle{Ecg-icovidnet: Interpretable ai model to identify changes in the ecg signals of post-covid subjects}.
\bjtitle{Computers in Biology and Medicine}
\bvolume{146},
\bfpage{105540}
(\byear{2022})
\end{barticle}
\endbibitem

\bibitem[\protect\citeauthoryear{Zubair et~al.}{2023}]{zubair2023deep}
\begin{botherref}
\oauthor{\bsnm{Zubair}, \binits{M.}},
\oauthor{\bsnm{Woo}, \binits{S.}},
\oauthor{\bsnm{Lim}, \binits{S.}},
\oauthor{\bsnm{Kim}, \binits{D.}}:
Deep representation learning with sample generation and augmented attention module for imbalanced ecg classification.
IEEE Journal of Biomedical and Health Informatics
(2023)
\end{botherref}
\endbibitem

\bibitem[\protect\citeauthoryear{Tao et~al.}{2022}]{tao2022ecg}
\begin{barticle}
\bauthor{\bsnm{Tao}, \binits{Y.}},
\bauthor{\bsnm{Li}, \binits{Z.}},
\bauthor{\bsnm{Gu}, \binits{C.}},
\bauthor{\bsnm{Jiang}, \binits{B.}},
\bauthor{\bsnm{Zhang}, \binits{Y.}}:
\batitle{Ecg-based expert-knowledge attention network to tachyarrhythmia recognition}.
\bjtitle{Biomedical Signal Processing and Control}
\bvolume{76},
\bfpage{103649}
(\byear{2022})
\end{barticle}
\endbibitem

\bibitem[\protect\citeauthoryear{Rosa and Yang}{2019}]{rosa2019flexible}
\begin{barticle}
\bauthor{\bsnm{Rosa}, \binits{B.M.}},
\bauthor{\bsnm{Yang}, \binits{G.Z.}}:
\batitle{A flexible wearable device for measurement of cardiac, electrodermal, and motion parameters in mental healthcare applications}.
\bjtitle{IEEE Journal of Biomedical and Health Informatics}
\bvolume{23}(\bissue{6}),
\bfpage{2276}--\blpage{2285}
(\byear{2019})
\end{barticle}
\endbibitem

\bibitem[\protect\citeauthoryear{Rosa et~al.}{2019}]{rosa2019nfc}
\begin{barticle}
\bauthor{\bsnm{Rosa}, \binits{B.M.G.}},
\bauthor{\bsnm{Anastasova-Ivanova}, \binits{S.}},
\bauthor{\bsnm{Yang}, \binits{G.Z.}}:
\batitle{Nfc-powered flexible chest patch for fast assessment of cardiac, hemodynamic, and endocrine parameters}.
\bjtitle{IEEE Transactions on Biomedical Circuits and Systems}
\bvolume{13}(\bissue{6}),
\bfpage{1603}--\blpage{1614}
(\byear{2019})
\end{barticle}
\endbibitem

\bibitem[\protect\citeauthoryear{Mazaheri and Khodadadi}{2020}]{mazaheri2020heart}
\begin{barticle}
\bauthor{\bsnm{Mazaheri}, \binits{V.}},
\bauthor{\bsnm{Khodadadi}, \binits{H.}}:
\batitle{Heart arrhythmia diagnosis based on the combination of morphological, frequency and nonlinear features of ecg signals and metaheuristic feature selection algorithm}.
\bjtitle{Expert Systems with Applications}
\bvolume{161},
\bfpage{113697}
(\byear{2020})
\end{barticle}
\endbibitem

\bibitem[\protect\citeauthoryear{Li et~al.}{2022}]{li2022umap}
\begin{barticle}
\bauthor{\bsnm{Li}, \binits{M.}},
\bauthor{\bsnm{Si}, \binits{Y.}},
\bauthor{\bsnm{Yang}, \binits{W.}},
\bauthor{\bsnm{Yu}, \binits{Y.}}:
\batitle{Et-umap integration feature for ecg biometrics using stacking}.
\bjtitle{Biomedical Signal Processing and Control}
\bvolume{71},
\bfpage{103159}
(\byear{2022})
\end{barticle}
\endbibitem

\bibitem[\protect\citeauthoryear{Safdar et~al.}{2023}]{safdar2023exploring}
\begin{botherref}
\oauthor{\bsnm{Safdar}, \binits{M.F.}},
\oauthor{\bsnm{Nowak}, \binits{R.M.}},
\oauthor{\bsnm{Pa{\l}ka}, \binits{P.}}:
Exploring artificial intelligence algorithms for electrocardiogram (ecg) signal analysis: A comprehensive review.
Computers in Biology and Medicine,
107908
(2023)
\end{botherref}
\endbibitem

\bibitem[\protect\citeauthoryear{Hong et~al.}{2020}]{hong2020cardioid}
\begin{barticle}
\bauthor{\bsnm{Hong}, \binits{S.}},
\bauthor{\bsnm{Wang}, \binits{C.}},
\bauthor{\bsnm{Fu}, \binits{Z.}}:
\batitle{Cardioid: learning to identification from electrocardiogram data}.
\bjtitle{Neurocomputing}
\bvolume{412},
\bfpage{11}--\blpage{18}
(\byear{2020})
\end{barticle}
\endbibitem

\bibitem[\protect\citeauthoryear{Kumar et~al.}{2023}]{kumar2023fuzz}
\begin{barticle}
\bauthor{\bsnm{Kumar}, \binits{S.}},
\bauthor{\bsnm{Mallik}, \binits{A.}},
\bauthor{\bsnm{Kumar}, \binits{A.}},
\bauthor{\bsnm{Del~Ser}, \binits{J.}},
\bauthor{\bsnm{Yang}, \binits{G.}}:
\batitle{Fuzz-clustnet: Coupled fuzzy clustering and deep neural networks for arrhythmia detection from ecg signals}.
\bjtitle{Computers in Biology and Medicine}
\bvolume{153},
\bfpage{106511}
(\byear{2023})
\end{barticle}
\endbibitem

\bibitem[\protect\citeauthoryear{Ma et~al.}{2020}]{ma2020electrocardiogram}
\begin{barticle}
\bauthor{\bsnm{Ma}, \binits{Y.}},
\bauthor{\bsnm{Sun}, \binits{S.}},
\bauthor{\bsnm{Zhang}, \binits{M.}},
\bauthor{\bsnm{Guo}, \binits{D.}},
\bauthor{\bsnm{Liu}, \binits{A.R.}},
\bauthor{\bsnm{Wei}, \binits{Y.}},
\bauthor{\bsnm{Peng}, \binits{C.-K.}}:
\batitle{Electrocardiogram-based sleep analysis for sleep apnea screening and diagnosis}.
\bjtitle{Sleep and Breathing}
\bvolume{24},
\bfpage{231}--\blpage{240}
(\byear{2020})
\end{barticle}
\endbibitem

\bibitem[\protect\citeauthoryear{Rosa and Yang}{2019}]{rosa2019low}
\begin{bchapter}
\bauthor{\bsnm{Rosa}, \binits{B.M.}},
\bauthor{\bsnm{Yang}, \binits{G.Z.}}:
\bctitle{A low-powered capacitive device for detection of heart beat and cardiovascular parameters}.
In: \bbtitle{2019 IEEE Biomedical Circuits and Systems Conference (BioCAS)},
pp. \bfpage{1}--\blpage{4}
(\byear{2019}).
\bcomment{IEEE}
\end{bchapter}
\endbibitem

\bibitem[\protect\citeauthoryear{Ginsburg et~al.}{2024}]{ginsburg2024key}
\begin{barticle}
\bauthor{\bsnm{Ginsburg}, \binits{G.S.}},
\bauthor{\bsnm{Picard}, \binits{R.W.}},
\bauthor{\bsnm{Friend}, \binits{S.H.}}:
\batitle{Key issues as wearable digital health technologies enter clinical care}.
\bjtitle{New England Journal of Medicine}
\bvolume{390}(\bissue{12}),
\bfpage{1118}--\blpage{1127}
(\byear{2024})
\end{barticle}
\endbibitem

\bibitem[\protect\citeauthoryear{Thirunavukarasu et~al.}{2023}]{thirunavukarasu2023large}
\begin{barticle}
\bauthor{\bsnm{Thirunavukarasu}, \binits{A.J.}},
\bauthor{\bsnm{Ting}, \binits{D.S.J.}},
\bauthor{\bsnm{Elangovan}, \binits{K.}},
\bauthor{\bsnm{Gutierrez}, \binits{L.}},
\bauthor{\bsnm{Tan}, \binits{T.F.}},
\bauthor{\bsnm{Ting}, \binits{D.S.W.}}:
\batitle{Large language models in medicine}.
\bjtitle{Nature medicine}
\bvolume{29}(\bissue{8}),
\bfpage{1930}--\blpage{1940}
(\byear{2023})
\end{barticle}
\endbibitem

\bibitem[\protect\citeauthoryear{Rashidinejad et~al.}{2020}]{rashidinejad2020patient}
\begin{barticle}
\bauthor{\bsnm{Rashidinejad}, \binits{P.}},
\bauthor{\bsnm{Hu}, \binits{X.}},
\bauthor{\bsnm{Russell}, \binits{S.}}:
\batitle{Patient-adaptable intracranial pressure morphology analysis using a probabilistic model-based approach}.
\bjtitle{Physiological measurement}
\bvolume{41}(\bissue{10}),
\bfpage{104003}
(\byear{2020})
\end{barticle}
\endbibitem

\bibitem[\protect\citeauthoryear{Wu et~al.}{2018}]{wu2018personalizing}
\begin{bchapter}
\bauthor{\bsnm{Wu}, \binits{M.-H.}},
\bauthor{\bsnm{Chang}, \binits{E.J.}},
\bauthor{\bsnm{Chu}, \binits{T.-H.}}:
\bctitle{Personalizing a generic ecg heartbeat classification for arrhythmia detection: a deep learning approach}.
In: \bbtitle{2018 IEEE Conference on Multimedia Information Processing and Retrieval (MIPR)},
pp. \bfpage{92}--\blpage{99}
(\byear{2018}).
\bcomment{IEEE}
\end{bchapter}
\endbibitem

\bibitem[\protect\citeauthoryear{Banluesombatkul et~al.}{2020}]{banluesombatkul2020metasleeplearner}
\begin{barticle}
\bauthor{\bsnm{Banluesombatkul}, \binits{N.}},
\bauthor{\bsnm{Ouppaphan}, \binits{P.}},
\bauthor{\bsnm{Leelaarporn}, \binits{P.}},
\bauthor{\bsnm{Lakhan}, \binits{P.}},
\bauthor{\bsnm{Chaitusaney}, \binits{B.}},
\bauthor{\bsnm{Jaimchariyatam}, \binits{N.}},
\bauthor{\bsnm{Chuangsuwanich}, \binits{E.}},
\bauthor{\bsnm{Chen}, \binits{W.}},
\bauthor{\bsnm{Phan}, \binits{H.}},
\bauthor{\bsnm{Dilokthanakul}, \binits{N.}}, \betal:
\batitle{Metasleeplearner: A pilot study on fast adaptation of bio-signals-based sleep stage classifier to new individual subject using meta-learning}.
\bjtitle{IEEE Journal of Biomedical and Health Informatics}
\bvolume{25}(\bissue{6}),
\bfpage{1949}--\blpage{1963}
(\byear{2020})
\end{barticle}
\endbibitem

\bibitem[\protect\citeauthoryear{Hwang et~al.}{2021}]{hwang2021pbgan}
\begin{barticle}
\bauthor{\bsnm{Hwang}, \binits{D.Y.}},
\bauthor{\bsnm{Taha}, \binits{B.}},
\bauthor{\bsnm{Hatzinakos}, \binits{D.}}:
\batitle{Pbgan: Learning ppg representations from gan for time-stable and unique verification system}.
\bjtitle{IEEE Transactions on Information Forensics and Security}
\bvolume{16},
\bfpage{5124}--\blpage{5137}
(\byear{2021})
\end{barticle}
\endbibitem

\bibitem[\protect\citeauthoryear{Jin et~al.}{2020}]{jin2020novel}
\begin{barticle}
\bauthor{\bsnm{Jin}, \binits{Y.}},
\bauthor{\bsnm{Qin}, \binits{C.}},
\bauthor{\bsnm{Liu}, \binits{J.}},
\bauthor{\bsnm{Lin}, \binits{K.}},
\bauthor{\bsnm{Shi}, \binits{H.}},
\bauthor{\bsnm{Huang}, \binits{Y.}},
\bauthor{\bsnm{Liu}, \binits{C.}}:
\batitle{A novel domain adaptive residual network for automatic atrial fibrillation detection}.
\bjtitle{Knowledge-Based Systems}
\bvolume{203},
\bfpage{106122}
(\byear{2020})
\end{barticle}
\endbibitem

\end{thebibliography}

 \nocite{*}

\end{document}